\documentclass[a4paper,fleqn,usenatbib,usedcolumn]{mnras}

\usepackage{newtxtext,newtxmath}
 
\usepackage[T1]{fontenc}
\usepackage{ae,aecompl}

\usepackage{graphicx}	
\usepackage{ctable}

\defcitealias{Khrykin2019}{Paper~I}
\defcitealias{Worseck2021}{Paper~II}

\title[Quasar lifetime distribution]{The first measurement of the quasar lifetime distribution}

\author[Ilya S. Khrykin]{
Ilya S. Khrykin,$^{1}$\thanks{E-mail: ilya.khrykin@ipmu.jp}
Joseph F. Hennawi,$^{2,3}$
G\'abor Worseck,$^{4}$
Frederick B. Davies$^{5,6}$\\
$^{1}$Kavli Insitute for Physics and Mathematics of the Universe (WPI), UTIAS, The University of Tokyo, Kashiwa, Chiba 277-8583, Japan\\
$^{2}$Department of Physics, University of California, Santa Barbara, CA 93106-9530, USA\\
$^{3}$Leiden Observatory, Leiden University, P.O. Box 9513, 2300 RA, Leiden, The Netherlands\\
$^{4}$Institut f\"ur Physik und Astronomie, Universit\"at Potsdam, Karl-Liebknecht-Str.\ 24/25, D-14476 Potsdam, Germany\\
$^{5}$Max-Planck-Institut f\"ur Astronomie, K\"onigstuhl 17, D-69117 Heidelberg, Germany\\
$^{6}$Lawrence Berkeley National Laboratory, 1 Cyclotron Rd, Berkeley, CA 94720, USA\\
}

\date{Accepted XXX. Received YYY; in original form ZZZ}

\pubyear{2021}

\begin{document}
\label{firstpage}
\pagerange{\pageref{firstpage}--\pageref{lastpage}}
\maketitle

\begin{abstract}
Understanding the growth of the supermassive black holes (SMBH) powering luminous quasars, their co-evolution with host galaxies, and impact on the surrounding intergalactic medium (IGM) depends sensitively on the duration of quasar accretion episodes. Unfortunately, this time-scale, known as the quasar lifetime, $t_{\rm Q}$, is still uncertain by orders of magnitude ($t_{\rm Q} \simeq 0.01~{\rm Myr} - 1~{\rm Gyr} $). However, the extent of the \ion{He}{ii} Ly$\alpha$ proximity zones in the absorption spectra of $z_{\rm qso} \sim 3-4$ quasars constitutes a unique probe, providing sensitivity to lifetimes up to $\sim 30$~Myr. Our recent analysis of $22$ archival \emph{Hubble Space Telescope} \ion{He}{ii} proximity zone spectra reveals a surprisingly broad range of emission timescales, indicating that some quasars turned on $\lesssim 1$~Myr ago, whereas others have been shining for $\gtrsim 30$~Myr. Determining the underlying quasar lifetime distribution (QLD) from  proximity zone measurements is a challenging task owing to: 1) the limited sensitivity of individual measurements; 2) random sampling of the quasar light curves; 3) density fluctuations in the quasar environment; and 4) the inhomogeneous ionization state of \ion{He}{ii} in a reionizing IGM. We combine a semi-numerical \ion{He}{ii} reionization model, hydrodynamical simulations post-processed with ionizing radiative transfer, and a novel statistical framework to infer the QLD from an ensemble of proximity zone measurements.  Assuming a log-normal QLD, we infer a mean $\langle {\rm log}_{10}\left( t_{\rm Q} / {\rm Myr} \right)\rangle = 0.22^{+0.22}_{-0.25}$ and standard deviation $\sigma_{{\rm log}_{10}t_{\rm Q}} = 0.80^{+0.37}_{-0.27}$. Our results allow us to estimate the probability of detecting very young quasars with $t_{\rm Q} \leq 0.1$~Myr from their proximity zone sizes yielding $p \left( \leq 0.1~{\rm Myr}\right) = 0.19^{+0.11}_{-0.09}$, which is broadly consistent with recent determination at $z\sim 6$. 
\end{abstract}




\section{Introduction}
\label{sec:intro}

Observational efforts of the recent decades resulted in the discovery of dormant supermassive black holes (with typical masses in the range $M_{\rm BH} \simeq 10^9$--$10^{10} M_{\sun}$) in the centres of all nearby bulge-dominated galaxies \citep{Soltan1982, Yu2002, Kormendy2013}. It is believed that these are the remnants of the once powerful and luminous quasar phase that galaxies went through in their evolution. As quasars, these black holes played an important role in the thermal and ionization evolution of the intergalactic medium \citep{McQuinn2009, Compostella2013, Chardin2017, DAloisio2017, Khrykin2017, LaPlante2017, Puchwein2019, Kulkarni2019, Dayal2020}. In addition, many models of galaxy formation and evolution include various quasar feedback prescriptions that are believed to have a significant impact on their host galaxies, for instance, regulating the star-formation rate \citep{Hopkins2006, Choi2018, Habouzit2019}. However, despite their cosmological importance, many aspects of quasar evolution remain relatively unconstrained. 

For example, while it is widely accepted that quasars are powered by accretion on to their supermassive black holes, we do not yet fully understand the physical mechanisms responsible for ``powering up'' the accretion. Quasar activity is thought to be triggered by either major galaxy mergers \citep[e.g.][]{DiMatteo2005, Springel2005a,  DiMatteo2005, Hopkins2005a, Hopkins2005c, Hopkins2008, Capelo2015, Steinborn2018, Banados2019}, or by secular disc instabilities \citep[e.g.][]{Goodman2003,  Hopkins2010, Hopkins2011a, Novak2011, Bournaud2011, Gabor2013, Angles-alcazar2013, Angles-alcazar2017}, or by some combination of the two. Another puzzling discovery is the existence of SMBHs with masses up to $M_{\rm SMBH}\simeq 10^{10} M_{\sun}$ in quasars already at $z \simeq 6$--7 \citep{Fan2006, Mortlock2011, Venemans2013, DeRosa2014, Wu2015, Mazzucchelli2017, Banados2018, Shen2019, Wang2020, Yang2020a, Wang2021}. Current theoretical models struggle to explain such high masses existing already less than $\sim 1$~Gyr after the Big Bang, and require both very massive initial seeds and Eddington-limited accretion of matter on time-scales comparable to the Hubble time, or super-Eddington accretion \citep{Hopkins2009, Volonteri2010, Volonteri2012, Smith2017, Regan2019, Inayoshi2019, Davies2019}.

Distinguishing between different theoretical models of SMBH/quasar evolution requires knowledge of the characteristic time-scale over which accretion on to SMBHs occurs. Unfortunately, observations have not yet converged on a coherent picture for this time-scale, the so-called quasar lifetime $t_{\rm Q}$, with a range of uncertain estimates to date covering several orders of magnitude ($0.01~{\rm Myr} \lesssim t_{\rm Q} \lesssim 1~{\rm Gyr}$; see \citet{Martini2004} for a review). For instance, studies of quasar clustering \citep{Haiman2001, Martini2001} constrain the total integrated time that a galaxy hosts an active quasar, known as the quasar duty cycle $t_{\rm dc}$. Measurements at $z\simeq 2$--4 have yielded only weak constraints $t_{\rm dc} \simeq 1~{\rm Myr}-1~{\rm Gyr}$ \citep{Shen2007, White2008, Eftekharzadeh2015}, mostly due to the large uncertainties in how the quasars populate the dark matter haloes \citep{Shen2009, White2012, Conroy2013, Cen2015}. Comparable results (with comparable uncertainties) $t_{\rm dc} \simeq 10-100$~Myr also come from the comparison of the integrated quasar luminosity function to the present day SMBH number density \citep{Soltan1982, Yu2002}. Most importantly, these methods only constrain the integrated quasar lifetime, and not the duration of individual accretion episodes $t_{\rm Q}$, which theoretical models suggest could be much shorter \citep{Ciotti2001, Novak2011, Angles-alcazar2017, Angles2020}.

\begin{figure}
\centering
\includegraphics[width=\columnwidth]{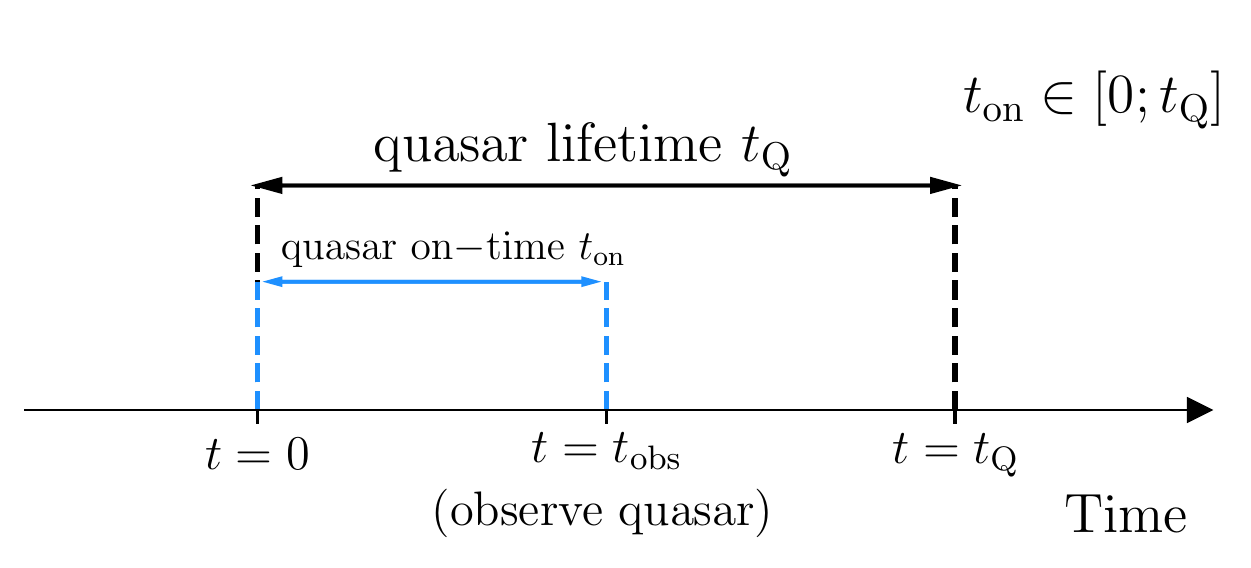}
\caption{\label{fig:ton} Illustration of the difference between the quasar lifetime $t_{\rm Q}$ and the lower limit on $t_{\rm Q}$, the quasar on-time $t_{\rm on}$, inferred from the analysis of individual quasar proximity zones.}
\end{figure}

The impact of the quasar's ionizing radiation on the surrounding IGM provides an independent observational probe of the quasar lifetime $t_{\rm Q}$. For example, the time delay between variations in the radiation field and the environment's response to these changes was used to argue for quasar variability on short $t_{\rm Q} \simeq 0.1$--$10$~Myr time-scales \citep{Hennawi2007, Schawinski2015, Schmidt2017,Eilers2017,Schmidt2018}. The sensitivity to the quasar lifetime also comes from the spatial extent of the regions around quasars where the ionization level of the IGM is enhanced by the quasar radiation. These regions, commonly known as line-of-sight proximity zones, were first detected in the \ion{H}{i} Ly$\alpha$ forest in the spectra of $z\simeq 2$--4 quasars 
\citep[e.g.][]{Carswell1982, Bajtlik1988, Dallaglio2008}.
The sensitivity of these proximity zones to the quasar lifetime arises
from the finite response time of the IGM to the variations in the quasar enhanced photoionization rate $\Gamma$. This {\it equilibration time-scale} $t_{\rm eq} \simeq \Gamma^{-1}$ sets the upper limit on the duration of episodic quasar activity that can be probed by their proximity zones \citep{Khrykin2016, Eilers2017, Davies2020a}. Moreover, the extent of the quasar proximity zones is actually sensitive only to the quasar {\it on-time}, $t_{\rm on} \leq t_{\rm Q}$, as illustrated in Fig.~\ref{fig:ton}. Imagine that the quasar light is detected by an observer at the time $t_{\rm obs}$, then the quasar on-time is defined such that the quasar accretion episode began at a time $t_{\rm obs}-t_{\rm on}$ in the past. However, this quasar episode may continue and end at a later time $t_{\rm Q}$, which can be recorded on Earth only if the observer could conduct observations of the same quasar in the future. For instance, the \ion{H}{i} proximity effect at $z \simeq 2$--4 provides only a lower limit on the quasar on-time $t_{\rm on}\geq t_{\rm eq}\simeq 0.03$~Myr, owing to the high post-reionization \ion{H}{i} background radiation ($\Gamma_{\rm HI} \simeq 10^{-12}\ {\rm s^{-1}}$; \citealp{Becker2013}). However, the situation is qualitatively different at $z \simeq 6$ where the mostly opaque \ion{H}{i} Ly$\alpha$ forest far away from quasars \citep{Fan2006, Eilers2018, Bosman2018, Yang2020b} provides a better contrast, and analysis of the quasar \ion{H}{i} proximity zones at $z\simeq 6$ has revealed several very young $t_{\rm on} \simeq 0.01$--$1$~Myr quasars \citep{Eilers2017,Eilers2020}. Moreover, an analysis of the composite \ion{H}{i} proximity zone profile of $15$ quasars at $z\gtrsim 6$ points to an average on-time of $t_{\rm on} \simeq 0.50^{+0.58}_{-0.25}$~Myr \citep{Morey2021}. This is consistent with finding newly turned on quasars with $t_{\rm on}\sim 0.01$-0.1~Myr approximately $\sim 1$-10~$\%$ of the time \citep{Eilers2020}.

An analogous proximity effect has also been detected in \ion{He}{ii} Ly$\alpha$ absorption spectra toward $z\sim 3$--4 quasars
\citep[e.g.][]{Hogan1997, Anderson1999, Syphers2014, Zheng2015},
where the contrast is similar to the \ion{H}{i} Ly$\alpha$ forest at $z\simeq6$. Recently,  \citet{Khrykin2016} showed that \ion{He}{ii} proximity zones probe longer and more interesting
time-scales up to $\simeq 30$~Myr owing to the \ion{He}{ii} ionizing background at $z\simeq 3$--4 being $3$ orders of magnitude lower 
\citep{Khrykin2016, Worseck2019, Makan2020}. These time-scales are comparable to the Salpeter time-scale ($t_{\rm S} \simeq 45$~Myr; \citealp{Salpeter1964}), the characteristic time for the SMBH evolution. 
In \citet[][hereafter \citetalias{Khrykin2019}]{Khrykin2019} we presented the first fully Bayesian statistical method to infer the quasar on-times of individual quasars from their \ion{He}{ii} proximity zones at $z \simeq 4$, and measured short lifetimes of order $1$~Myr for several quasars. We recently expanded our inference to the full sample of \ion{He}{ii}-transparent quasars, discovering a surprisingly broad distribution of individual on-times $t_{\rm on}$, ranging from $\lesssim 1$~Myr to $\gtrsim 30$~Myr \citep[][hereafter \citetalias{Worseck2021}]{Worseck2021}. The nature of such wide scatter is yet unknown, but it might indicate that the extent of the \ion{He}{ii} proximity zones samples a broad underlying distribution of quasar lifetimes.

\begin{figure*}
\includegraphics[width=\textwidth]{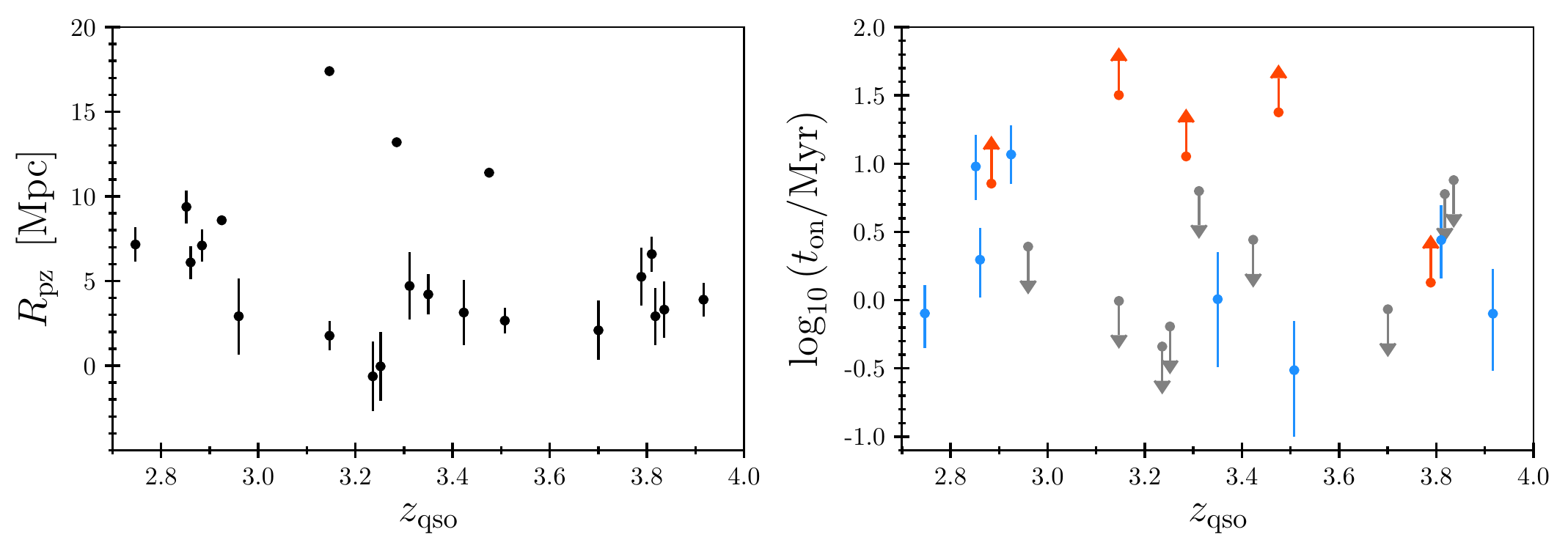}
\caption{\label{fig:rpz_data}
\textit{Left:} The sizes of the \ion{He}{ii} proximity zone in the spectra of $22$ quasars used in this work as a function of redshift. The error bars correspond to the measured redshift uncertainty (see Table~\ref{tab:table1} for details). \textit{Right:} Inferred values of quasar on-times from \citetalias{Khrykin2019} and \citetalias{Worseck2021}. The blue dots with error bars show the measurements, while the red and grey arrows indicate $1\sigma$ lower and upper limits, respectively.}
\end{figure*}

In this work, we introduce a fully Bayesian statistical algorithm to recover the intrinsic quasar lifetime distribution. We re-analyse the sample of \ion{He}{ii} proximity zones from \citetalias{Khrykin2019} and \citetalias{Worseck2021} with similar numerical modelling but with a modified Bayesian approach to measure for the first time the underlying quasar lifetime distribution.
In contrast to previous analyses of \ion{He}{ii} proximity zones \citep{Syphers2014, Zheng2015} that often ignored the apparent degeneracy between the quasar lifetime and the ionization state of the IGM \citep{Khrykin2016}, our algorithm instead takes this degeneracy into account and allows us to fully marginalize over the poorly constrained ionization state of \ion{He}{ii} in the IGM. Moreover, for individual quasars $t_{\rm on}$ is often poorly constrained owing to large errors in proximity zone sizes resulting from large quasar redshift errors \citepalias{Khrykin2019, Worseck2021}. Our Bayesian approach takes these errors into account and optimally combines these weak constraints from individual quasars to infer the underlying lifetime distribution.

This paper is organized in the following way. In Section~\ref{sec:data} we outline the main properties of the quasars in our sample. We summarize our hydrodynamical and radiative transfer simulations in Section~\ref{sec:sims}, and discuss their results in Section~\ref{sec:rpz_sims}. In Section~\ref{sec:stats} we describe our statistical algorithm for measuring the lifetime distribution and present the results of our inference from Markov Chain Monte Carlo (MCMC). We discuss our findings in Section~\ref{sec:disc} and conclude in Section~\ref{sec:conc}.

Throughout this work, we assume a flat $\Lambda$CDM cosmology with dimensionless Hubble constant $h=0.7$, $\Omega_m = 0.27$, $\Omega_b = 0.046$, $\sigma_8=0.8$, and $n_s = 0.96$,  the helium  mass fraction is $Y_{\rm He} = 0.24$, consistent with the latest {\it Planck} results \citep{Planck2018}. All quoted distances are in units of proper Mpc if not stated otherwise. Throughout this work we quote all quasar lifetimes and on-times in the form ${\rm log_{10}}\left(t /{\rm Myr} \right)$.

\section{Data sample}
\label{sec:data}

Our parent sample of 24 $2.7 \lesssim z_{\rm qso} \lesssim 3.9$ quasars  observed with Cosmic Origins Spectrograph \citep[COS;][]{Green2012} on board the \textit{Hubble Space Telescope} (\textit{HST}) has been described in \citetalias{Khrykin2019} and \citetalias{Worseck2021}, to which we refer the interested reader. As in our previous work, we exclude two quasars with formally negative proximity zone sizes due to associated absorption (HE~2347$-$4342) or a likely underestimated systemic redshift caused by an anomalously large blueshift of the \ion{C}{iv} emission line (HE2QS~J2354$-$2033). We summarize the main properties of the remaining $22$ quasars used in this work in Table~\ref{tab:table1}.

Similar to the study of the \ion{H}{i} proximity zones at $z\simeq 6$ \citep{Fan2006}, we identify the size of the \ion{He}{ii} proximity zone $R_{\rm pz}$ as the proper distance from the quasar location where smoothed \ion{He}{ii} transmission profile drops below the $10$~per cent threshold for the first time. We adopt a Gaussian filter with ${\rm FWHM} = 1$~proper Mpc and apply it to all quasar spectra in our sample at their respective redshifts. The second to last column in Table~\ref{tab:table1} shows the measured proximity zone sizes from \citetalias{Khrykin2019} and \citetalias{Worseck2021}. We adopt these measurements for this work and show the \ion{He}{ii} proximity zone sizes as a function of redshift in the left panel of Fig.~\ref{fig:rpz_data}.

Quasar redshift uncertainty constitutes the dominant source of error in estimating the size of the observed \ion{He}{ii} proximity zones. All quasar redshifts in our sample were estimated from the available emission lines in rest-frame low- and medium-resolution UV-optical spectra. Four quasars in our sample have the smallest redshift uncertainty due to the detected narrow [\ion{O}{iii}] emission lines with a velocity precision of $44\ {\rm km\ s^{-1}}$. Among the remaining quasars with no [\ion{O}{iii}] detection, $6$ have broad \ion{Mg}{ii} lines with corresponding uncertainties $\Delta v = 273\ {\rm km\ s^{-1}}$; $3$ quasars have neither [\ion{O}{iii}] nor \ion{Mg}{ii} lines covered, but $H\beta$ is present in the $K$-band spectra, with uncertainties of $\Delta v = 400\ {\rm km\ s^{-1}}$. Finally, the remaining $7$ quasars do not have high-quality near-IR spectra, and their redshifts are determined from the \ion{C}{iv} line in the optical band. We take into account the luminosity dependent blueshift of the \ion{C}{iv} line \citep[the Baldwin effect;][]{Baldwin1977}, and find the high uncertainty of $\Delta v = 656\ {\rm km\ s^{-1}}$ calibrated against the quasars with known redshift \citepalias{Khrykin2019, Worseck2021}. We provide the identified redshifts and associated redshift uncertainties in Table~\ref{tab:table1}. We refer the interested reader to a detailed description of the redshift measurement procedure provided in \citetalias{Worseck2021}.

\ctable[caption = {Sample of $22$ quasars used in this work. From left to right, the columns show: quasar name, quasar redshift with measured uncertainty, base-$10$ logarithm of the \ion{He}{ii} ionizing photon production rate, measured size of the \ion{He}{ii} proximity zone with corresponding $1\sigma$ uncertainty, and the inferred quasar on-time $t_{\rm on}$ \citepalias{Khrykin2019, Worseck2021}.}, width = \columnwidth, center, doinside = \scriptsize]{X c c c c}{

\label{tab:table1}

 }{
\hline\hline  \\
Quasar & $z_{\rm qso}$  & $\log_{10}Q$ & $R_{\rm pz}$ & $t_{\rm on}$ \\ &  & ${\rm s}^{-1}$ & ${\rm Mpc}$ &  ${\rm Myr}$\\  [1.ex]  \hline
\\ [3ex] 
HS~1700$+$6416     & $2.7472 \pm 0.0034 $ & $57.34$ & $7.16  \pm 1.01$  & $0.80^{+0.50}_{-0.35}$\\
HS~1024$+$1849     & $2.8521 \pm 0.0035 $ & $56.62$ & $9.38  \pm 0.97$  & $9.53^{+6.83}_{-4.08}$\\
Q~1602$+$576       & $2.8608 \pm 0.0035 $ & $56.80$ & $6.10  \pm 0.97$  & $1.98^{+1.41}_{-0.94}$\\
PC~0058$+$0215     & $2.8842 \pm 0.0035 $ & $56.19$ & $7.10  \pm 0.97$  & $>7.24$\\
SDSS~J0936$+$2927  & $2.9248 \pm 0.0006 $ & $56.49$ & $8.59  \pm 0.15$  & $11.62^{+7.37}_{-4.57}$\\
SDSS~J0818$+$4908  & $2.9598 \pm 0.0087 $ & $56.38$ & $2.92  \pm 2.25$  & $<2.42$\\
HE2QS~J2157$+$2330 & $3.1465 \pm 0.0006 $ & $56.72$ & $17.40 \pm 0.14$  & $>31.84$\\
SDSS~J1237$+$0126  & $3.1467 \pm 0.0038 $ & $56.27$ & $1.77  \pm 0.87$  & $<1.01$\\
HE2QS~J1706$+$5904 & $3.2518 \pm 0.0093 $ & $56.27$ & $-0.04 \pm 2.03$
& $<0.66$\\
HE2QS~J2149$-$0859 & $3.2358 \pm 0.0093 $ &	$56.34$ & $-0.63 \pm 2.04$ & $< 0.46$ \\
Q~0302$-$003       & $3.2850 \pm 0.0006 $ & $56.89$ & $13.20 \pm 0.13$  & $>11.36$\\
HE2QS~J0233$-$0149 & $3.3115 \pm 0.0094 $ & $56.47$ & $4.71  \pm 1.98$  & $<6.24$\\
HS~0911$+$4809     & $3.3500 \pm 0.0058 $ & $56.74$ & $4.21  \pm 1.19 $ & $1.01^{+1.29}_{-0.69}$\\
HE2QS~J0916$+$2408 & $3.4231 \pm 0.0097 $ & $56.45$ & $3.14  \pm 1.91$  & $<2.77$\\
SDSS~J1253$+$6817  & $3.4753 \pm 0.0007 $ & $56.48$ & $11.40 \pm 0.12$  & $>23.55$\\
SDSS~J2346$-$0016  & $3.5076 \pm 0.0041 $ & $56.79$ & $2.66  \pm 0.77$  & $0.31^{+0.41}_{-0.21}$\\
HE2QS~J2311$-$1417 & $3.7003 \pm 0.0103 $ & $56.66$ & $1.94  \pm 1.72$  & $< 0.86$ \\
SDSS~J1137$+$6237  & $3.7886 \pm 0.0105 $ & $56.19$ & $4.92  \pm 1.68$  & $>1.34$ \\
HE2QS~J1630$+$0435 & $3.8101 \pm 0.0064 $ & $56.92$ & $8.43  \pm 1.02$  & $2.77^{+2.27}_{-1.33}$\\
SDSS~J1614$+$4859  & $3.8175 \pm 0.0105 $ & $56.14$ & $2.72  \pm 1.66$  & $<5.98$ \\ 
SDSS~J1711$+$6052  & $3.8358 \pm 0.0106 $ & $56.19$ & $2.97  \pm 1.65$  & $<7.48$\\
SDSS~J1319$+$5202  & $3.9166 \pm 0.0066 $ & $56.82$ & $3.62  \pm 0.98$  & $0.80^{+0.88}_{-0.50}$\\
[2ex]
\hline
}

In \citetalias{Khrykin2019} we introduced a Bayesian method for measuring the on-times of individual quasars based on a statistical comparison between the observed and simulated sizes of the \ion{He}{ii} proximity zones. We successfully applied this method to measure on-times of individual \ion{He}{ii}-transparent quasars at $2.7 \lesssim z \lesssim 3.9$ \citepalias{Khrykin2019, Worseck2021}. We show the results
in the right panel of Fig.~\ref{fig:rpz_data}. Note that these values are measured individual on-times $t_{\rm on}$, whereas in what follows we focus on the intrinsic quasar lifetime distribution.

\section{Simulations}
\label{sec:sims}

In this study, we utilize the set of simulations previously used in \citetalias{Khrykin2019} and \citetalias{Worseck2021}, although with modifications. In what follows, we briefly outline the main parameters of our models, and refer the interested reader to our previous papers for a more detailed description of the simulations.

The impact of quasar radiation on the IGM is modelled through two stages. First, we utilize the outputs of cosmological hydrodynamical simulations performed with the Gadget-3 code ($2\times 512^3$ particles and $L_{\rm box}=25\ h^{-1}$ cMpc; \citealp{springel2005}) and extract $1000$ realistic 1D density, velocity and temperature distributions (to which we refer as {\it skewers}) from $z_{\rm sim} = \left[ 3.1, 3.7, 3.9 \right]$ snapshots\footnote{We used $z_{\rm sim}=3.7$ and $z_{\rm sim}=3.9$ snapshots for the analysis of quasars at $z_{\rm qso} > 3.5$ \citepalias[see][]{Khrykin2019}, and $z_{\rm sim}=3.1$ snapshot for quasars at $z_{\rm qso} \leq 3.5$ \citepalias[see][]{Worseck2021} } using periodic boundary conditions\footnote{Starting from the location of the quasars, we create skewers by casting rays through the simulation volume at random angles, and traversing the box multiple times. We use the periodic boundary conditions to wrap a skewer through the box
along the chosen direction.}. The resulting skewers have a physical length of $160$~cMpc, and a spatial resolution of ${\rm d}r = 0.012$~comoving Mpc (${\rm d}v \approx 1.0~{\rm km\ s}^{-1}$). 

Next, we post-process the extracted skewers with a custom 1D radiative transfer algorithm based on the ${\rm C}^2$-Ray code \citep{mellema2006}, which tracks the evolution of \ion{H}{i}, \ion{He}{ii}, $e^{-}$, and gas temperature \citep{Khrykin2016,Khrykin2017}. For each quasar in our sample (see Table~\ref{tab:table1}) we create a set of radiative transfer models depending on the quasar redshift $z_{\rm qso}$, quasar photon production rate $Q_{\rm 4Ry}$, assumed quasar on-time $t_{\rm on}$, and initial fraction of \ion{He}{ii} $x_{\rm HeII,0}$ (set by the \ion{He}{ii} ionizing background prior to quasar activity). We compute the $Q_{\rm 4Ry}$ given the observed $i$-band magnitudes of the quasars in our sample (see \citetalias{Khrykin2019} and \citetalias{Worseck2021} for details), and together with observed $z_{\rm qso}$ fix them for each quasar that we simulate. In order to better capture the cosmological evolution of the density field between the redshift of the simulation output $z_{\rm sim}$ and the redshift of quasars $z_{\rm qso}$ in our sample, we rescale the simulated densities $\rho_{\rm sim}$ in every pixel along the skewers as 

\begin{figure}
\centering
\includegraphics[width=\columnwidth]{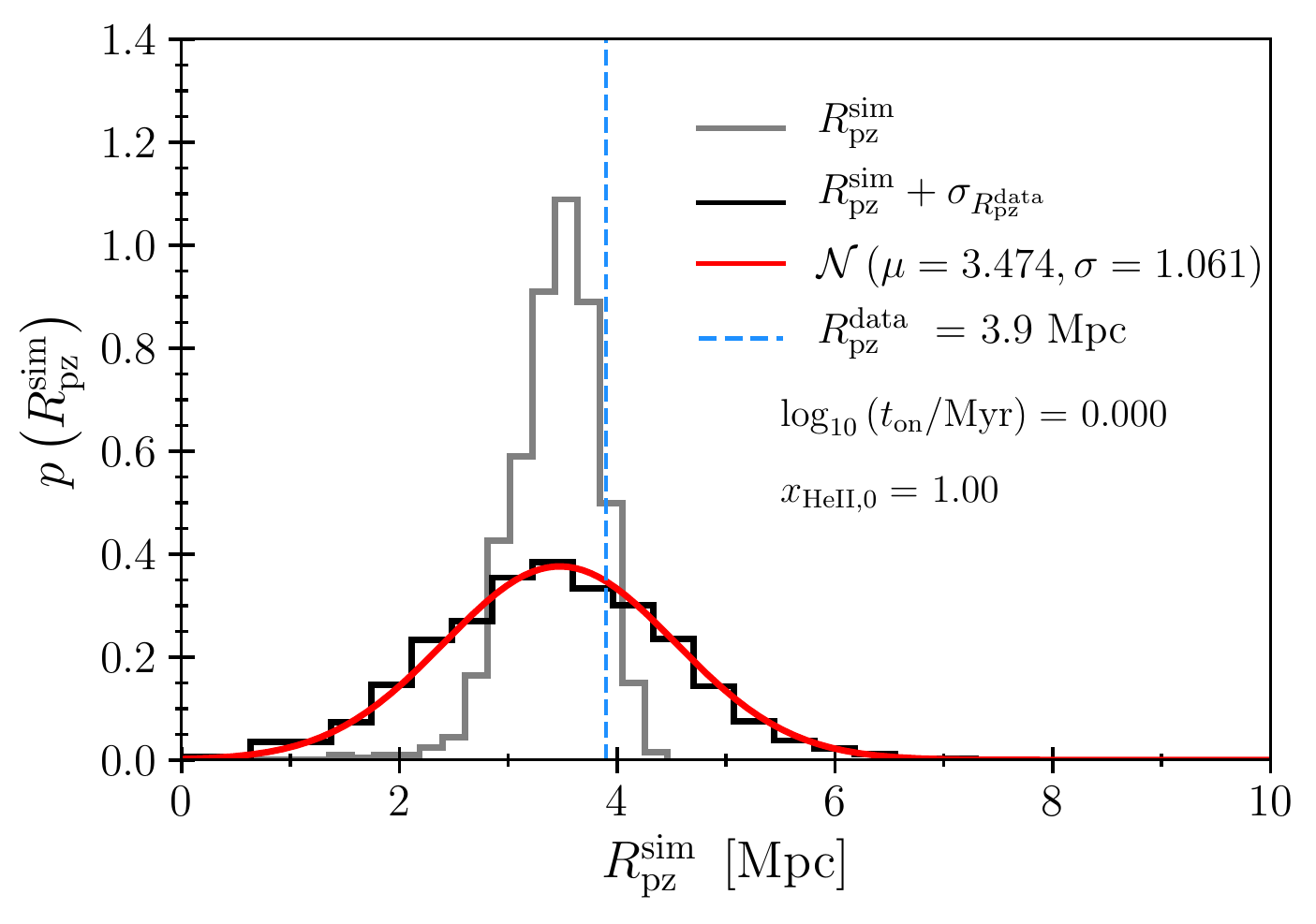}
\caption{\label{fig:rpz_dist}
Size distribution of the simulated \ion{He}{ii} proximity zones for quasar SDSS~J1319$+$5202 assuming a quasar on-time $1$~Myr and initial \ion{He}{ii} fraction $x_{\rm HeII,0} = 1.00$. The grey histogram illustrates the $R_{\rm pz}^{\rm sim}$ distribution without redshift uncertainties, while the black one incorporates them (see discussion in Section~\ref{sec:rpz_sims}). The blue dashed line marks the size of the observed \ion{He}{ii} proximity zone $R_{\rm pz}^{\rm data}$ for the quasar in question, while the red curve illustrates the Gaussian fit to the distribution.}
\end{figure}

\begin{equation}
    \rho \left( z_{\rm qso} \right) = \rho_{\rm sim} \times \frac{\left( 1 + z_{\rm qso}\right)^3}{\left(1 + z_{\rm sim}\right)^3}
\end{equation}

\begin{figure*}
\includegraphics[width=\textwidth]{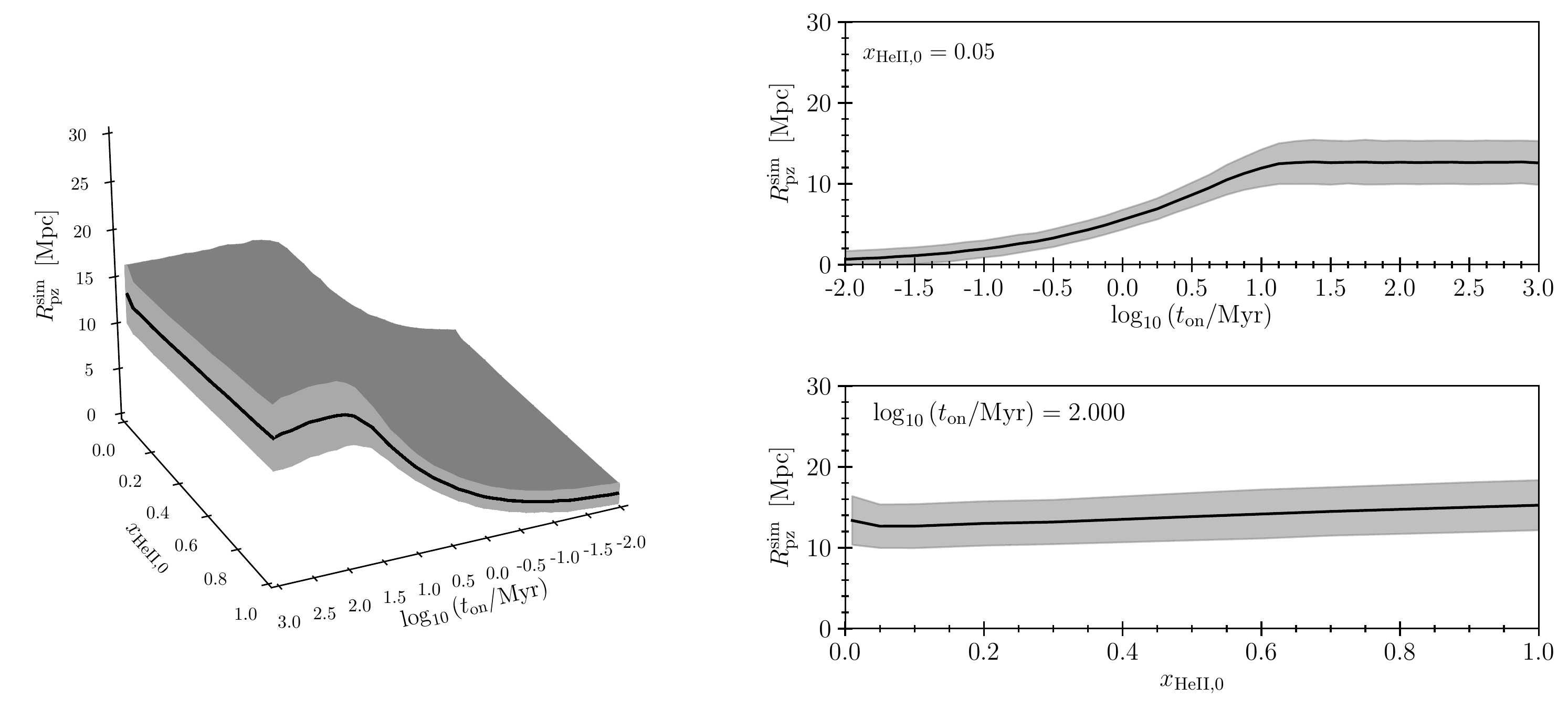}
\caption{\label{fig:rpz_evol}
Left: Evolution of the simulated \ion{He}{ii} proximity zone sizes $R_{\rm pz}^{\rm sim}$ with incorporated redshift uncertainties for quasar SDSS~J1319$+$5202 as a function of quasar on-time and initial \ion{He}{ii} fraction. The black solid curve shows the mean of the simulated $R_{\rm pz}^{\rm sim}$ distributions, while the shaded area illustrates the 1$\sigma$ variation. Right: $2$D slices through the $3$D plot on the left. The upper panel shows the evolution of the $R_{\rm pz}^{\rm sim}$ distribution parameters as a function of quasar on-time with fixed $x_{\rm HeII,0} = 0.05$, while the bottom panel shows the same evolution but as a function of initial \ion{He}{ii} fraction with fixed quasar on-time ${\rm log}_{10} \left( t_{\rm on} / {\rm Myr} \right) = 2.000$.}
\end{figure*}

In contrast to the previously used simulations \citepalias{Khrykin2019, Worseck2021}, we extended the upper limit of the logarithmically spaced quasar on-time values from ${\rm log_{10}}\left(t_{\rm on} /{\rm Myr} \right) = 2.000$ to ${\rm log_{10}}\left(t_{\rm on} /{\rm Myr} \right) = 3.000$, which results in the range ${\rm log_{10}}\left(t_{\rm on} /{\rm Myr} \right) = \left[ -2.000, 3.000 \right]$ with step $\Delta {\rm log_{10}}\left(t_{\rm on} /{\rm Myr} \right) = 0.125$. Similar to the analysis in \citetalias{Worseck2021}, we extended our original grid of models \citepalias{Khrykin2019} along the axis of the initial \ion{He}{ii} fraction by including additional models with $x_{\rm HeII,0} = 0.01$. This allows us to better capture the ionization state of the IGM at redshifts $z_{\rm qso} \lesssim 3.0$, in the tail end of \ion{He}{ii} reionization, where the \ion{He}{II} fractions are of order of one per cent \citep{Worseck2019}. Therefore, the initial \ion{He}{ii} fraction in our models can take one of the following values $x_{\rm HeII,0}\in \{ 0.01, 0.05, 0.10, 0.20, 0.30, 0.50, 0.60, 0.70, 0.90, 1.00 \}$. 

This results in a grid of $410$ radiative transfer models per quasar ($41$ values of quasar on-time and $10$ values of initial \ion{He}{ii} fraction) with $1000$ \ion{He}{ii} Ly$\alpha$ transmission spectra per model, $9020$ models in total for $22$ quasars.

\section{Simulated proximity zone sizes }
\label{sec:rpz_sims}

We smooth the simulated \ion{He}{ii} Ly$\alpha$ transmission spectra and measure the sizes of the proximity zones $R_{\rm pz}^{\rm sim}$ following the same algorithm applied to observational data (see Section~\ref{sec:data}), which results in the distribution of $1000$ $R_{\rm pz}^{\rm sim}$ for each model. Note, because the adopted smoothing scale is larger than observed spectral resolution and spans several pixels, the resulting extent of the \ion{He}{ii} proximity zones only weakly depends on the observational effects like the signal-to-noise ratio, or spectral resolution \citepalias[see discussion in Appendix A of][]{Worseck2021}. Therefore, we do not include these effects in our numerical simulations. Fig.~\ref{fig:rpz_dist} shows an example of the distribution of \ion{He}{ii} proximity zone sizes for one model of the quasar SDSS~J1319$+$5202 (see Table~\ref{tab:table1}). Because redshift uncertainty introduces significant error into the proximity zone size measurement, we take it into account by adding to the $R_{\rm pz}^{\rm sim}$ distribution random errors drawn from the Gaussian distribution with standard deviation equal to the observed redshift uncertainty, which for the case of SDSS~J1319$+$5202 is
$\sigma_z = 0.0066$ corresponding to $\sigma_{R_{\rm pz}^{\rm data}} = 0.98$~Mpc ($\sigma_v = 400\ {\rm km\ s^{-1}}$; see Table~\ref{tab:table1}).

Each $R_{\rm pz}^{\rm sim}$ distribution is well described by a Gaussian, which is apparent from the red line in Fig.~\ref{fig:rpz_dist}. For each quasar and each of the 410 models parametrized by $\{\log_{10}\left( t_{\rm on} / {\rm Myr} \right), x_{\rm HeII,0}\}$, we fit the resulting $R_{\rm pz}^{\rm sim}$ (with incorporated redshift uncertainties) with a Gaussian by determining its mean and standard deviation. This allows us to easily track the evolution of the distribution of proximity zone sizes as a function of the model parameters, i.e, quasar on-time $\log_{10} \left( t_{\rm on} / {\rm Myr} \right) $ and initial \ion{He}{ii} fraction $x_{\rm HeII,0}$ (see Section~\ref{sec:sims}). We illustrate an example of this evolution for the models of the quasar SDSS~J1319$+$5202 in Fig.~\ref{fig:rpz_evol}. It is apparent that the spatial extent of the \ion{He}{ii} proximity zones is sensitive to the quasar on-time until it is comparable to the equilibration time $t_{\rm eq}$ \citep{Khrykin2016}, at which point the $R_{\rm pz}^{\rm sim}$ saturates and sensitivity to the on-time diminishes as $R_{\rm pz}^{\rm sim}$ stops growing with increasing on-time. On the other hand, as \citetalias{Khrykin2019} showed, the size of the \ion{He}{ii} proximity zone is largely insensitive to the initial \ion{He}{ii} fraction. This arises from the competition between the decrease in the level of IGM transmission for large initial $x_{\rm HeII,0}$ fractions, and the increase in transmission due to the thermal proximity effect for high $x_{\rm HeII,0}$ fractions \citep{Khrykin2017}. Moreover, the very definition of the proximity zone size $R_{\rm pz}$, i.e., the location where transmission drops below the $10$ per cent threshold, reduces sensitivity to the \ion{He}{ii} background (which sets the $x_{\rm HeII,0}$). This is because at $10$ per cent transmission levels the quasar radiation still dominates the total photoionization rate, and the effect of the \ion{He}{ii} ionizing background on the transmission becomes prominent at larger distances corresponding to the transmission levels much lower than $10$ per cent.

We use bivariate spline interpolation to predict the mean and standard deviation of the $R_{\rm pz}^{\rm sim}$ distributions for values of ${\rm log}_{10} \left( t_{\rm on} / {\rm Myr} \right) $ and $x_{\rm HeII,0}$ at locations between our simulated parameter grid points. We use the dependence of the $R_{\rm pz}^{\rm sim}$ distributions on the model parameters in the likelihood calculations in the next section.

\section{Statistical Inference}
\label{sec:stats}

In this section, we describe our algorithm for measuring the parameters of the quasar lifetime distribution. In what follows, we assume that all inferred on-times of individual quasars (see Table~\ref{tab:table1}) are
drawn from the underlying statistical distribution. Moreover, we assume that the quasar lifetime distribution (to which we further refer to as QLD) is described by a log-normal distribution 

\begin{equation}
\label{eq:qld}
    p\left( t_{\rm Q} \right) = \frac{1}{ t_{\rm Q} } \frac{{\rm log_{10}}e}{\sigma \sqrt{2\pi}} \times {\rm exp} \left[ - \frac{ \left( {\rm log_{10}}\left( t_{\rm Q} / {\rm Myr} \right) - \mu \right)^2 }{ 2 \sigma^2 } \right],
\end{equation}
where $\mu = \langle {\rm log_{10}}\left(t_{\rm Q} /{\rm Myr} \right)  \rangle $ and $\sigma = \sigma_{{\rm log}_{10} t_{\rm Q} } $ are the mean and standard deviation of the distribution, respectively.  In what follows we show how we can infer these parameters. 

\subsection{Likelihood calculations}
\label{sec:like}

In order to estimate the parameters of the QLD we need to determine the Bayesian likelihood function for the observed proximity zone sizes  $\mathcal{L}_{\rm qso}\left( R_{{\rm pz},i}^{\rm data} | \mu , \sigma \right) $ for each quasar in our sample.

We begin by constructing a grid of parameter values on which the likelihood is estimated. For the mean of the QLD $\mu$ we adopt the range $\mu = \langle {\rm log_{10}}\left(t_{\rm Q} /{\rm Myr} \right) \rangle = \left[ -2.000, 2.000 \right] $ with step $\Delta \mu = 0.125$, while for the standard deviation of the QLD $\sigma$ we choose $ \sigma = \sigma_{ {\rm log}_{10} t_{\rm Q} } = \left[ 0.01, 3.0 \right]$~dex, with step $\Delta \sigma = 0.1$. We obtain the values of the individual likelihood $\mathcal{L}_{\rm qso}\left( R_{{\rm pz},i}^{\rm data} | \mu , \sigma \right)$ for a single quasar in our sample at each point on the parameter grid as follows:

\begin{enumerate}
\label{eq:list}
    \item Given the values of the mean $\mu$ and standard deviation $\sigma$ of the QLD we randomly draw $1000$ quasar lifetime values ${\rm log_{10}}\left(t_{\rm Q} /{\rm Myr} \right)$ from the resulting log-normal distribution given by eq.~(\ref{eq:qld}).
    
    \item In order to account for the fact that our observations measure the quasar on-times, $t_{\rm on}$, which represent a random sampling of the quasars light-bulb light curve, we construct a distribution of $t_{\rm on}$ times for each value of ${\rm log_{10}}\left(t_{\rm Q} /{\rm Myr} \right)$ drawn from the QLD. For that we assume a uniform distribution of on-times $ t_{\rm on} \sim \mathcal{U}\left( 0, t_{\rm Q} \right)$ (see Fig.~\ref{fig:ton}). We draw $10$ $t_{\rm on }$ values for each $t_{\rm Q}$ value, which results in $10000$ $t_{\rm on}$ values per location on the QLD parameter grid ($\mu, \sigma$). 
    
    \item For each $t_{\rm on}$ value we need to find the corresponding distribution of the proximity zone sizes in our simulations. The $R_{\rm pz}^{\rm sim}$ distributions, however, also depend on the value of initial \ion{He}{ii} fraction (see Fig.~\ref{fig:rpz_evol}), i.e., $R_{\rm pz}^{\rm sim}\left( {\rm log_{10}}\left(t_{\rm on}/{\rm Myr}\right), x_{\rm HeII,0} \right)$. In order to take this into account we use the semi-numerical model of the \ion{He}{ii} ionizing background, $\Gamma_{\rm HeII}$, from \citet{Davies2017}, which provides $1000$ realizations of $\Gamma_{\rm HeII}$ at each quasar redshift in our sample. Assuming a density equal to the mean density of the Universe at the redshift in question and that the IGM is in ionization equilibrium with the predicted $\Gamma_{\rm HeII}$, we calculate the distribution of $1000$ $x_{\rm HeII}$ values at this redshift. For each value of the on-time we then randomly draw a value of $x_{\rm HeII,0}$ from this distribution, which results in $10000$ $\{ {\rm log_{10}}\left(t_{\rm on}/{\rm Myr}\right), x_{\rm HeII,0} \}$ pairs per evaluation of the QLD. By doing so, we effectively perform a Monte-Carlo sampling of the \ion{He}{ii} fraction and marginalize out the unknown ionization state of the IGM. Finally, we use bivariate spline interpolation to find the mean and standard deviation of the corresponding Gaussian $R_{\rm pz}^{\rm sim}$ distribution at each parameter location $\{ {\rm log_{10}}\left(t_{\rm on}/{\rm Myr}\right), x_{\rm HeII,0} \}$ (see Section~\ref{sec:rpz_sims}).
    
    \item Using the Gaussian fits at these $10000$  $\{ {\rm log_{10}}\left(t_{\rm on}/{\rm Myr}\right), x_{\rm HeII,0} \}$ locations, we draw $500$ $R_{{\rm pz}}^{\rm sim}$ samples from the Gaussian fit at each parameter value and concatenate them into one 
    combined  distribution of the \ion{He}{ii} proximity zone sizes corresponding to all $\{ {\rm log_{10}}\left(t_{\rm on}/{\rm Myr}\right), x_{\rm HeII,0} \}$ values. This procedure has then effectively marginalized over the stochastic relationship between $t_{\rm on}$ and $t_{\rm Q}$, as well as stochastic distribution of $x_{\rm HeII,0}$ in the IGM at each redshift. 
    
    \item Finally, we apply Kernel Density Estimation (KDE) to find the continuous probability density function (PDF) of this combined $R_{\rm pz}^{\rm sim}$ distribution. We calculate the value of the likelihood by evaluating the KDE PDF at the observed value of the \ion{He}{ii} proximity zone size $R_{\rm pz}^{\rm data}$ for the quasar in question (see Table~\ref{tab:table1});
    
    \item we repeat steps (i)-(v) for each combination of the QLD parameters. 

\end{enumerate}

This procedure results in $990$ determinations of the likelihood function on the $\{ \mu , \sigma \}$ parameter grid for each quasar. Next, we obtain a joint likelihood for all quasars in our sample simply by taking the product of the individual (independent) likelihoods, calculated following the procedure stated above

\begin{equation}
\label{eq:joint_likelihood}
    \mathcal{L_{\rm joint}} = f\left( R_{\rm pz}^{\rm data} \right) =  \prod_{i = 0}^{N_{\rm qso}} \mathcal{L}_{\rm qso}\left( R_{{\rm pz},i}^{\rm data} | \mu , \sigma \right),
\end{equation}
where $N_{\rm qso} = 22$ is the number of the quasars in the sample.

Finally, we use a bivariate spline to interpolate the $\mathcal{L_{\rm joint}}$ given eq.~(\ref{eq:joint_likelihood}) for any combination of parameter values between the grid points in our parameter space. We now can sample this joint likelihood with MCMC to estimate the posterior distributions of the QLD mean $\mu $ and standard deviation $\sigma $.
However, first, we need to determine the priors for each parameter.

\subsection{Choice of priors}
\label{sec:priors}

We assume a flat logarithmic prior on the mean of the QLD, i.e., $\mu = \langle {\rm log}_{10}\left( t_{\rm Q} / {\rm Myr}\right) \rangle = \left[ -2.000, 2.000 \right]$. As discussed in \citetalias{Khrykin2019}, the choice of this range is motivated by the current estimates of the quasar lifetime \citep[see][for a review]{Martini2004}. The lower limit is set by the existence of the line-of-sight proximity effect in the \ion{H}{i} Ly$\alpha$ forest, because it requires the quasar to shine at least for $t_{\rm eq} \simeq 0.03$~Myr \citep{Bajtlik1988, Khrykin2016}. On the other hand, the upper limit is chosen to lie in the upper range of estimates provided by the analysis of the quasar clustering and models of the SMBH growth \citep{Shen2007, White2008, Eftekharzadeh2015}. 

\begin{figure}
\centering
\includegraphics[width=\columnwidth]{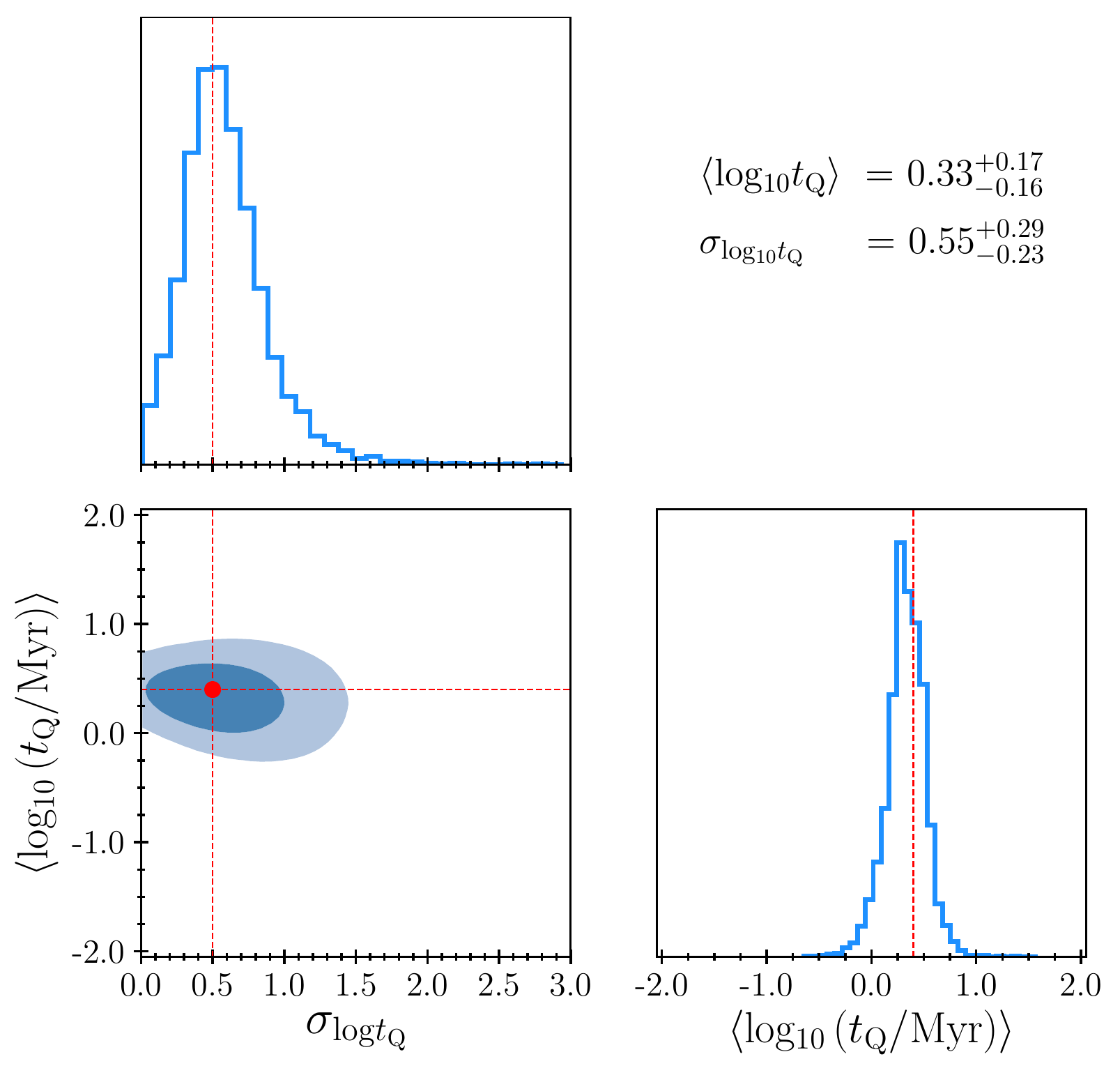}
\caption{\label{fig:mcmc_mock_sample}
Results of the MCMC inference on the mock quasar sample.
The bottom left panel shows the $2$D contours, while to top left and bottom right panels illustrate the constrained marginalized posterior probabilities of the mean $\mu$ and standard deviation $\sigma$ of the mock QLD. The red dot shows the values of the input QLD parameters}
\end{figure}

For the standard deviation of the QLD we also adopt a flat uniform prior in range $\sigma = \left[ 0.01, 3.0 \right]$~dex. Although other observations do not provide much guidance on what range should be adopted, it is motivated by the large scatter found in the present estimates of the quasar lifetime and by the variety of methods. 

\begin{figure*}
\centering
\includegraphics[width=\textwidth]{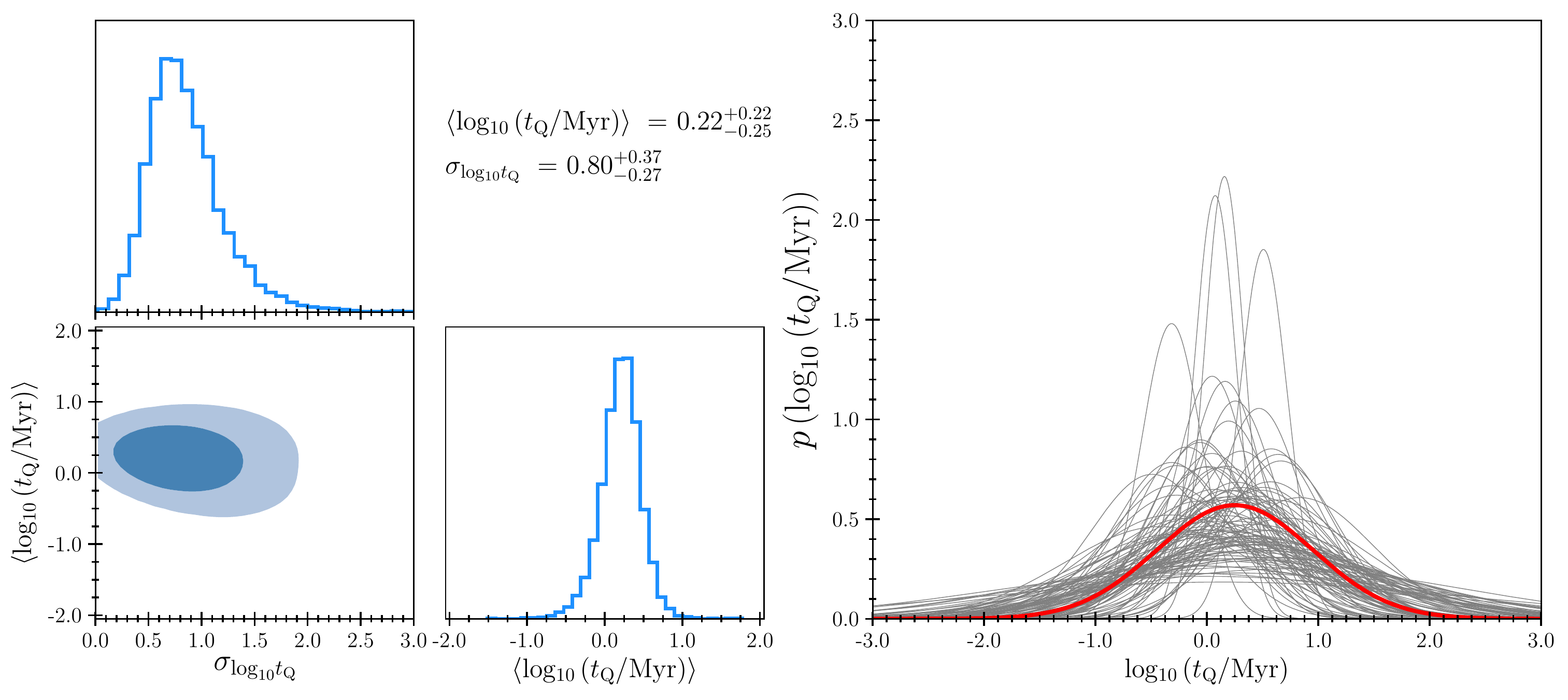}
\caption{\label{fig:mcmc}
 Results of the MCMC inference on the observed sample of quasars in Table~\ref{tab:table1}. Similar to Fig.~\ref{fig:mcmc_mock_sample}, the left-hand side panels show inferred $2$D contours and marginalized posterior probabilities of the QLD parameters. The right-hand side panel illustrates $100$ realizations of the QLD based on random draws of the mean $\mu$ and standard deviation $\sigma$ from the MCMC chains. The solid red curve corresponds to the maximum likelihood values of mean and standard deviation.}
\end{figure*}

However, we note that combined with the prior on the mean of the log-normal distribution, this range allows quasar lifetimes longer than $100$~Myr, which can be problematic for our modelling strategy. Namely, the post-processing radiative transfer approach that we use assumes that cosmic structures do not evolve over the time-scales of the calculations. We also neglect cooling of the gas due to cosmological expansion and recombination \citep{Hui1997, Khrykin2017}. Both assumptions start to break down on time-scales comparable to the Hubble time, which at $z \simeq 4$ is $\simeq 1.6$~Gyr, and, in principle, can lead to changes in the structure of the underlying gas distribution around the quasar, affecting the spatial extent of the simulated proximity zones. We acknowledge these limitations, but nevertheless choose to extend our simulation grid up to ${\rm log_{10}} \left( t_{\rm on} / {\rm Myr} \right) = 3.000$ (see Section~\ref{sec:sims}) and allow the wide prior range on $\sigma_{\rm log_{10} t_{\rm Q}}$. Our reasoning for this is that \ion{He}{ii} proximity zones are anyway not sensitive to lifetimes $t_{\rm on} \gtrsim t_{\rm eq} \simeq 30$~Myr at $z \simeq 3$--4 \citep{Khrykin2016}, as is apparent from Fig.~\ref{fig:rpz_evol} where $R_{\rm pz}$ saturates at $t_{\rm on} \sim t_{\rm eq}$ and does not significantly increase all the way up to $t_{\rm on} \sim 1$~Gyr. This lack of sensitivity implies that our results will not be sensitive to limitations of our model for $t_{\rm on} \gtrsim 100$~Myr. 

\subsection{MCMC results}
\label{sec:mcmc}

Given the expression for the joint likelihood in eq.~(\ref{eq:joint_likelihood}), and the interpolation procedure which allows us to evaluate this likelihood at any point in our parameter space, we can now sample this likelihood using MCMC and obtain the posterior probability distribution for our model parameters. In what follows we use the publicly available affine-invariant MCMC sampling algorithm {\it emcee} \citep{Foreman-Mackey2013}. 

\subsection{Analysis of the mock dataset}
\label{sec:mcmc_mock}

First, to test the accuracy of our inference procedure, we apply it to a mock sample of $N_\mathrm{qso}=22$ quasars (the same number as in our observed sample listed in Table~\ref{tab:table1}). In order to create the mock sample of \ion{He}{ii} proximity zones we perform the following steps.

We begin by choosing the mock QLD parameters. We choose the mean $\mu = 0.40$ and standard deviation $\sigma=0.50$ to describe the mock QLD, in anticipation of the results in the next section. From this distribution, we randomly draw $N_{\rm qso}=22$ values of the quasar lifetime ${\rm log}_{10}\left( t_{\rm Q} / {\rm Myr} \right)$. Using the uniform probability distribution for the on-times given the quasar lifetime (see discussion in Section~\ref{sec:like}), we draw one $t_{\rm on}$ value for each ${\rm log}_{10}\left( t_{\rm Q} / {\rm Myr} \right)$. This results in the sample of $N_{\rm qso} = 22$ on-times, i.e., one for each mock quasar. Further, we assume that quasars in our mock sample map to the real quasars in our dataset, and assign them a corresponding redshift $z_{\rm qso}$ and photon production rate consistent with Table~\ref{tab:table1}. We use these redshifts to assign a value of the initial \ion{He}{ii} fraction $x_{\rm HeII,0}$ to each quasar by randomly drawing a value from the corresponding $x_{\rm HeII}$ distribution at redshift $z_{\rm qso}$ similar to our procedure in Section~\ref{sec:like}. Next, for each quasar in this mock sample, characterized by the values of on-time and \ion{He}{ii} fraction, we find a corresponding distribution of the \ion{He}{ii} proximity zone sizes $R_{\rm pz}^{\rm sim}\left( {\rm log_{10}}\left(t_{\rm on}/{\rm Myr}\right), x_{\rm HeII,0} \right)$ from our radiative transfer simulations. We randomly draw one value of $R_{\rm pz}^{\rm mock}$ for each quasar from the corresponding $R_{\rm pz}^{\rm sim}$ distribution, which now corresponds to the ``observed" sample of the mock \ion{He}{ii} proximity zone sizes. In order to reproduce the accuracy of the observed \ion{He}{ii} proximity zones we also add the redshift associated uncertainties to each $R_{\rm pz}^{\rm mock}$, for which we adopt the values of $\sigma_{\rm R_{\rm PZ}}$ listed in Table~\ref{tab:table1}. 

We then proceed to calculate the joint likelihood of the mock sample of quasars following the discussion in Section~\ref{sec:like}, and sample it with MCMC. Fig.~\ref{fig:mcmc_mock_sample} illustrates the results of this inference.  The dark blue (light blue) contours correspond to the $68$ per cent ($95$ per cent) confidence regions, while the marginalized posterior PDFs of the mean and standard deviation of the QLD are shown by the histograms. The red dot shows the values of input QLD parameters.
We quote the $50^{\rm th}$ percentiles of the 1D marginalized posterior probability distributions as the measured values of the inferred parameters, whereas their uncertainties are derived from the $16^{\rm th}$ and $84^{\rm th}$ percentiles.

It is apparent from Fig.~\ref{fig:mcmc_mock_sample} that our inference algorithm successfully recovers the input parameters of the mock QLD with $\approx 0.1-0.3$~dex precision for both parameters. We therefore proceed to apply our method to the observed sample of quasars in the next Section.

\subsection{Analysis of the observed dataset}
\label{sec:mcmc_real}

The corner plot in the left panel of Fig.~\ref{fig:mcmc} shows the result of the MCMC inference of the QLD parameters $\{ \mu , \sigma \}$ for the $20$ quasars in our observed sample (see Table~\ref{tab:table1}).

We measure the mean of the QLD $\mu = 0.22^{+0.22}_{-0.25}$, and the QLD standard deviation $\sigma = 0.80^{+0.37}_{-0.27}$ from the corresponding marginalized posterior distributions shown by the histograms in the corner plot of Fig.~\ref{fig:mcmc}. We further illustrate our findings in the right panel of Fig.~\ref{fig:mcmc}, where $200$ realizations of the QLD based on random draws from MCMC samples of the mean and standard deviation are shown. The solid red curve corresponds to the maximum likelihood values of the mean ($\mu \simeq 0.25$) and standard deviation ($\sigma \simeq 0.70$) of the QLD. 

It is apparent from the left panels of Fig.~\ref{fig:mcmc} that while our MCMC algorithm is able to constrain the QLD parameters to the expected $\approx 0.2-0.3$~dex precision, the values for the standard deviation $\sigma \simeq 0.0$ is not completely excluded by our inference. We note that such small values seem to be implausible given the quite broad distribution of the inferred quasar on-times (see Fig.~\ref{fig:rpz_data}). On the other hand, it might also indicate that our sensitivity to the standard deviation of the QLD is limited by  the stochasticity in the sampling of quasar on-times $t_{\rm on}$ from the corresponding log-normal distribution of quasar lifetimes $t_{\rm Q}$.

\section{Discussion}
\label{sec:disc}

In what follows, we examine how the redshift uncertainties affect the constraining power of our analysis, and discuss our findings in the context of previous measurements of the quasar on-times and lifetimes, as well as the significance of our results for SMBH evolution models.

\subsection{Effect of the redshift uncertainties}
\label{sec:redshift}

In \citetalias{Khrykin2019} we showed that large uncertainties in the estimated quasar systemic redshifts, $z_{\rm qso}$, modify the distributions of the simulated \ion{He}{ii} proximity zone sizes, making them wider (see the comparison between the grey and black histograms in Fig.~\ref{fig:rpz_dist}), thus significantly weakening individual constraints on the quasar on-times $t_{\rm on}$. Therefore, for individual measurements, it would be ideal to obtain systemic redshifts from sub-millimeter observations of CO and/or [\ion{C}{ii}] $158\ \mu {\rm m}$ lines, decreasing the uncertainty to $\Delta v \simeq 100\ {\rm km\ s^{-1}}$ \citep{Eilers2020}.  In principle, the redshift uncertainties might also degrade the constraints on the QLD parameters. In order to quantify the significance of redshift precision on the results of our inference we perform the following test. 

\begin{figure}
\centering
\includegraphics[width=\columnwidth]{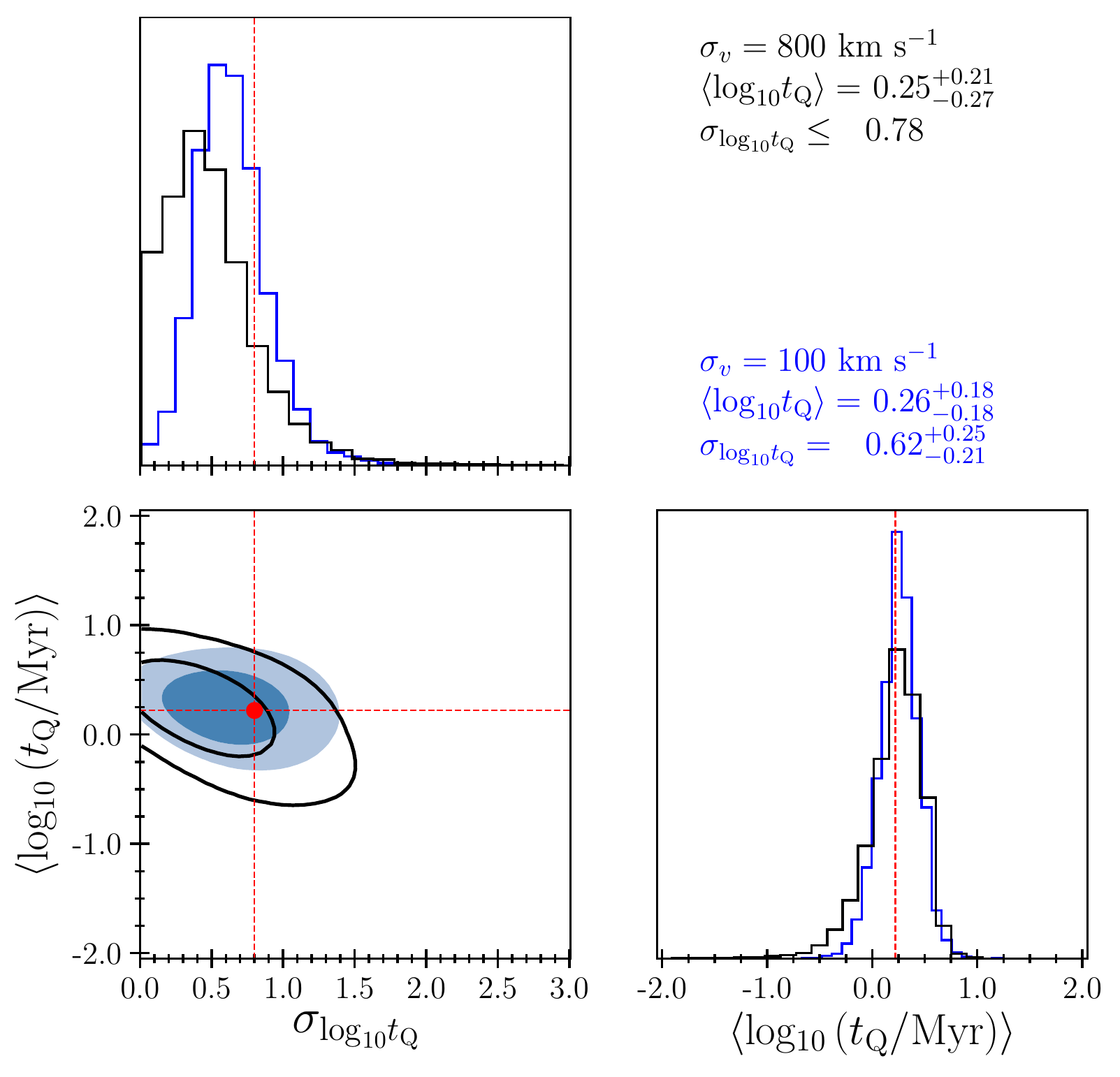}
\caption{\label{fig:mcmc_redshift}
Results of the MCMC inference on the mock quasars sample. Similar to Fig.~\ref{fig:mcmc_mock_sample}, the bottom left panel shows the $2$D contours, while to top left and bottom right panels illustrate the constrained marginalized posterior probabilities of the QLD parameters. The filled contours (and blue posterior histograms) show the result for the ``ideal" case of redshift uncertainties ($\sigma_v = 100~{\rm km\ s^{-1}}$), while the black curves correspond to the case with ``bad" redshift uncertainties ($\sigma_v = 800~{\rm km\ s^{-1}}$). The red dot indicates the input values of the QLD parameters.}
\end{figure}

Analogous to the discussion in Section~\ref{sec:mcmc_mock}, we begin by creating a mock sample of $22$ quasars with the same photon production rates and redshifts as our observed catalog (see Table~\ref{tab:table1}). For the values of the QLD model that we simulate, we choose the median best-fitting values of the QLD parameters inferred in Section~\ref{sec:mcmc_real}, i.e, $\mu = \log_{10}\left( {t_{\rm Q}} / {\rm Myr} \right) = 0.22$ and $\sigma_{\log_{10}t_{\rm Q}} = 0.80$. In what follows we assume that the same grid of radiative transfer models used for observed quasar sample (see Section~\ref{sec:sims}) corresponds to this mock dataset. 

Similarly to the discussion in Section~\ref{sec:rpz_sims} (see also \citetalias{Khrykin2019} and \citetalias{Worseck2021}), the impact of the redshift uncertainties is modelled by adding random Gaussian distributed proximity zone size errors with standard deviation $\sigma_{R_{\rm pz}}$ to the simulated distributions of the \ion{He}{ii} proximity zone sizes of each quasar. In order to test the sensitivity of our inference algorithm to the redshift uncertainty, we consider two cases: 1) ``bad" - the redshift uncertainties for each quasar in the mock sample are comparable to the biggest uncertainties in our observed sample (see discussion in Section~\ref{sec:data}) and correspond to the velocity precision $\sigma_v = 800\ {\rm km\ s^{-1}}$, and 2) ``ideal", with the redshift uncertainties corresponding to $\sigma_v = 100\ {\rm km\ s^{-1}}$, which is the precision expected from future sub-millimeter observations (it is also comparable to the values inferred from the analysis of the [\ion{O}{iii}] line used to measure systemic redshifts of several quasars in our sample). The value of $\sigma_{ R_{\rm pz} }$ for each mock quasar is then found via $\sigma_{ R_{\rm pz}} = \sigma_v / H\left( z_{\rm qso} \right)$, where $H\left( z_{\rm qso} \right)$ is the Hubble parameter at the redshift of a mock quasar.

\begin{figure*}
\centering
\includegraphics[width=\textwidth]{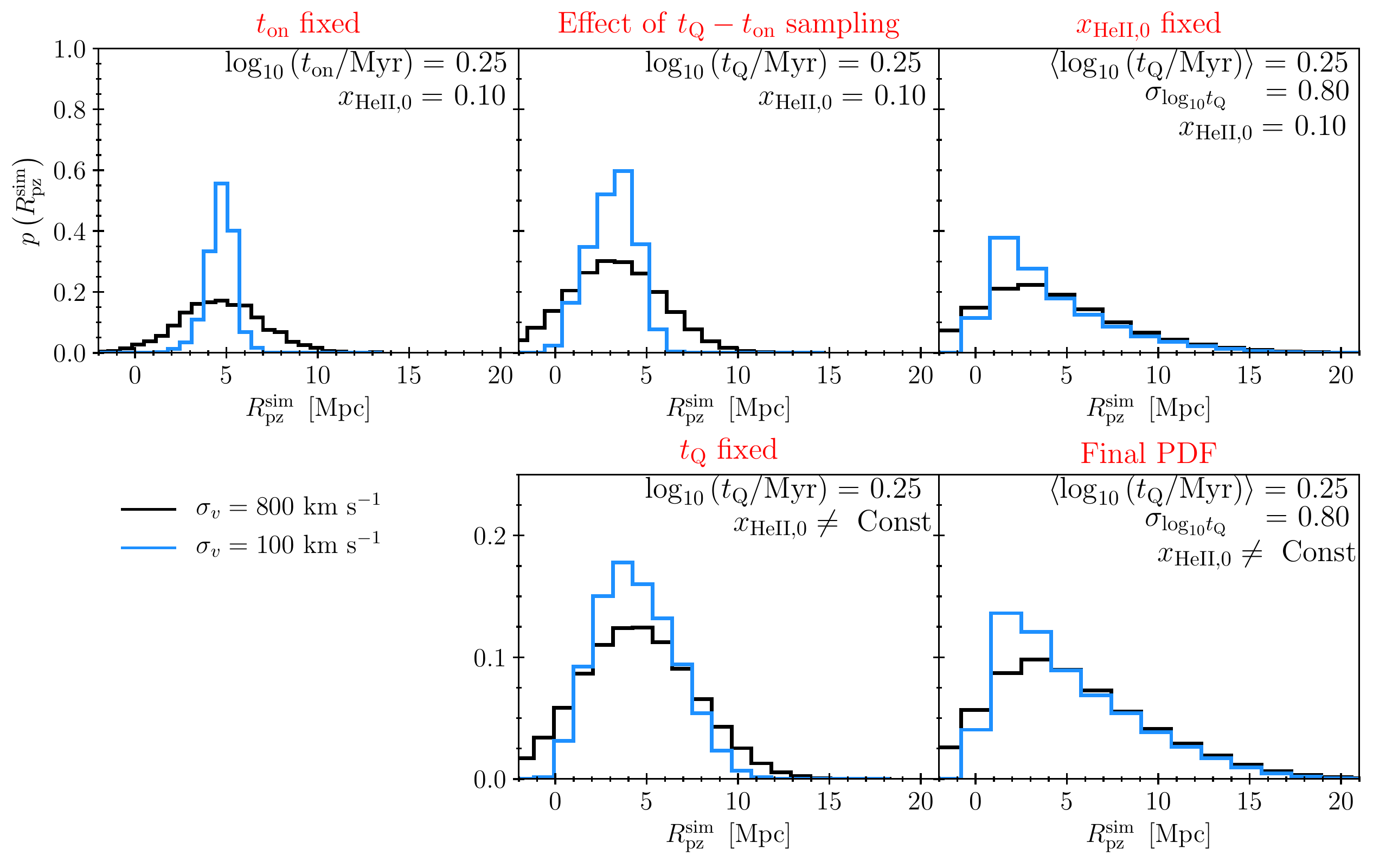}
\caption{\label{fig:stochast}
Different sources of stochasticity in our inference algorithm. The black and blue histograms illustrate two cases of redshift uncertainties corresponding to velocity precision of $\sigma_v = 800\ {\rm km\ s^{-1}}$ (``bad") and $\sigma_v = 100\ {\rm km\ s^{-1}}$ (``ideal"), respectively. The panels in the top row illustrate: {\it left:} the effect of the redshift uncertainties on the simulated \ion{He}{ii} proximity zone sizes distribution in the models of a single quasar HE2QS~J0233$-$0149; {\it middle:} the effect of sampling the $t_{\rm Q}-t_{\rm on}$ relation on the joint $R_{\rm pz}$ distribution (see Section~\ref{sec:like} for details); {\it right:} the effect of sampling the assumed log-normal distribution of quasar lifetimes on joint $R_{\rm pz}$ distribution for a single realization of the QLD. The value of \ion{He}{ii} fraction is fixed at $x_{\rm HeII,0} = 0.10$. The bottom {\it middle} panel shows the combined effects of sampling the $t_{\rm Q}-t_{\rm on}$ relation and variations in the assumed \ion{He}{ii} fraction on the joint $R_{\rm pz}$ distribution. Finally, the bottom {\it right} panel illustrates the same effect as shown in the top right panel, but combined with variations in the \ion{He}{ii} fraction. 
}
\end{figure*}

Finally, following the discussion in Section~\ref{sec:like}, we perform the calculations on these noisy distributions and compute the joint likelihood function given by eq.~(\ref{eq:joint_likelihood}) for these two cases. We sample each of these likelihoods with MCMC and compare the results in Fig.~\ref{fig:mcmc_redshift}. It is apparent from the constraints in Fig.~\ref{fig:mcmc_redshift} that the difference between the two cases is not very significant, implying a weak dependence on the precision with which we infer the QLD parameters on the level of redshift uncertainty. Although this result might be surprising in light of the individual measurements of $t_{\rm on}$ \citepalias{Khrykin2019, Worseck2021} where redshift errors play a large role, this weak dependence on redshift precision arises from the interplay of several factors. Our inference algorithm presented in Section~\ref{sec:stats} is subject to several sources of stochasticity that directly influence the resulting sensitivity to the QLD parameters. We illustrate the effect of each of these sources in Fig.~\ref{fig:stochast} and elaborate on them separately in what follows. We note that for simplicity, the discussion that follows focuses on a single quasar, HE2QS~J0233$-$0149, located close to the mean redshift of our sample ($z_{\rm qso} = 3.3115$; see Table~\ref{tab:table1}). 

First, consider the distribution of simulated \ion{He}{ii} proximity zone sizes in one model of HE2QS~J0233$-$0149, illustrated in the top left panel of Fig.~\ref{fig:stochast}, where the quasar on-time is fixed at ${\rm log_{10}}\left( t_{\rm on} / {\rm Myr} \right)  = 0.25$ and $x_{\rm HeII,0} = 0.10$. It is apparent that the $R_{\rm pz}^{\rm sim}$ distributions are significantly different depending on the magnitude of the redshift uncertainty, i.e, the distribution is much wider in the ``bad" case characterized by the velocity precision $\sigma_v = 800\ {\rm km\ s^{-1}}$. In \citetalias{Khrykin2019} we illustrated that incorporating larger redshift uncertainties decreases the constraining power of individual proximity zone measurements. Therefore, individual measurements of quasar on-times depend significantly on the accuracy of the quasar redshifts \citepalias[][\citealp{Eilers2020}]{Khrykin2019,Worseck2021}.

However, the situation is different if one tries to infer the quasar lifetime distribution. As we showed in Section~\ref{sec:like}, the measured individual quasar on-times represent random sampling of the quasar light bulb light curve. Mathematically, this sampling is described by the uniform distribution $t_{\rm on} \sim \mathcal{U}\left[ 0, t_{\rm Q} \right]$. The middle panel in the top row of Fig.~\ref{fig:stochast} labelled "Effect of $t_{\rm Q}-t_{\rm on}$ sampling" shows the distributions of \ion{He}{ii} proximity zone sizes constructed by 
sampling the single quasar lifetime of $ {\rm log_{10}} \left( t_{\rm Q} / {\rm Myr} \right) = 0.25$ for the "bad" and "ideal" redshift scenarios.  Similar to the discussion in Section~\ref{sec:like}, we randomly draw $1000$ values of $t_{\rm on}$ from $\mathcal{U}\left[ 0, t_{\rm Q} \right]$, and for each $t_{\rm on}$ value and fixed value of $x_{\rm HeII,0} = 0.10$, we find the corresponding $R_{\rm pz}^{\rm sim}$ distribution from our radiative transfer model grid. We then draw $500$ $R_{\rm pz}^{\rm sim}$ values from each Gaussian fit and combine the resulting proximity zone sizes in one joint distribution shown in the middle panel. It is clear that $t_{\rm Q} - t_{\rm on}$ sampling broadens the resulting $R_{\rm pz}^{\rm sim}$ distribution irrespective of the assumed redshift uncertainty, in contrast with case of fixed $t_{\rm on}$ (illustrated in the top left panel of Fig.~\ref{fig:stochast}) where the differences between "bad" and "ideal" redshifts is larger. 

The foregoing example only captures the stochasticity due to $t_{\rm on}$ sampling from a single quasar lifetime. In reality, there is an underlying QLD, which, as we assumed in Section~\ref{sec:stats} is described by a log-normal distribution (see eq.~(\ref{eq:qld})). The top right panel in Fig.~\ref{fig:stochast} demonstrates the combined effect of the QLD sampling and the previously described $t_{\rm Q} - t_{\rm on}$ sampling on the resulting joint $R_{\rm pz}^{\rm sim}$ distributions. For this exercise we choose one QLD realization with the mean $\mu = 0.25$ and standard deviation $\sigma = 0.80$, and keep the \ion{He}{ii} fraction fixed at $x_{\rm HeII,0} = 0.10$ for all models. As one can see, the  $R_{\rm pz}^{\rm sim}$ distributions flatten out and become less peaked, further reducing the difference between the two cases of redshift uncertainties. 

The situation is further aggravated when variations in the poorly known \ion{He}{ii} fraction $x_{\rm HeII,0}$ in the surrounding IGM are also taken into account. In the middle panel of the bottom row of Fig.~\ref{fig:stochast}, we show the same $t_{\rm Q} - t_{\rm on}$ sampling for a single value of quasar lifetime as before (top middle panel), but now including variations in $x_{\rm HeII,0}$. As described in Section~\ref{sec:like}, the value of $x_{\rm HeII,0}$ is randomly drawn from the \ion{He}{ii} fraction PDF from our semi-numerical model at the redshift of quasar in question. It is apparent that the $R_{\rm pz}^{\rm sim}$ distributions are significantly broadened when compared to the case of fixed \ion{He}{ii} fraction in the upper middle panel. The difference between two cases of redshift uncertainties is almost completely mitigated. 

Finally, the right panel in the bottom row of Fig.~\ref{fig:stochast}  illustrates the combined impact of all sources of stochasticity on the resulting joint $R_{\rm pz}^{\rm sim}$ distributions for the two cases of redshift error precision. It is evident that the two redshift uncertainty scenarios are basically indistinguishable. Consequently, the likelihood function evaluated from these distributions will be practically identical (see Section~\ref{sec:like} for details), which explains the large degree of overlap between the posterior probability distributions of the inferred QLD parameters for these two cases shown in Fig.~\ref{fig:mcmc_redshift}. 

According to the results presented in Fig.~\ref{fig:stochast}, we conclude that our inference algorithm only weakly depends on the exact quasar redshift uncertainties. This is because these redshift errors are swamped by a much larger scatter produced by the combination of several other sources of stochasticity (i.e. $t_{\rm Q}-t_{\rm on}$ sampling, the QLD itself, and fluctuations in $x_{\rm HeII,0}$). It is, therefore, possible to obtain tight constraints on the QLD parameters (see results in Section~\ref{sec:mcmc_real}) even using quasar samples with poorly determined redshifts. Note, however, that as one can see from the top left panel of Fig.~\ref{fig:stochast}, good precision on measured quasar redshifts is nevertheless required to accurately infer individual quasar on-times \citepalias[see also][]{Khrykin2019,Worseck2021}. This is especially important for the study of young quasars with exceptionally small proximity zones \citep{Eilers2020}.

\subsection{Comparison to individual measurements of quasar on-times and previous estimates of quasar lifetimes}
\label{sec:ton_comp}

In order to directly compare the results of our inference presented in Section~\ref{sec:mcmc_real} to the individual measurements of the on-times $t_{\rm on}$ from the quasar \ion{He}{ii} proximity zones \citepalias{Khrykin2019, Worseck2021}, we need to convert the inferred QLD into the distribution of on-times.

Assuming that the QLD is described by eq.~(\ref{eq:qld}), we derive an analytical formula for the $p\left( {\rm log}_{10}\left( t_{\rm on} / {\rm Myr} \right) \right)$ -- the probability density function for the quasar on-times ${\rm log}_{10}\left( t_{\rm on} / {\rm Myr} \right)$ as a function of the mean and standard deviation of the QLD -- given by
\begin{equation}
\label{eq:ansol}
    \begin{aligned}
    p\left( {\rm log}_{10} \left( t_{\rm on} / {\rm Myr} \right) \right) = \frac{2^{-\mu-1/2}~ 5^{-\mu}}{\sqrt{2}}{\rm exp}\left[ \frac{1}{2}\sigma^2 {\rm ln^2}10 \right] \cdot \\
    {\rm Erfc}\left[ \frac{{\rm log_{10}}\left(t_{\rm on}/{\rm Myr}\right) - \mu + \sigma^2{\rm ln}10}{\sigma\sqrt{2}} \right] \cdot {\rm ln}10~t_{\rm on},
    \end{aligned}
\end{equation}
where ${\rm Erfc}$ is the complementary error function, and $\mu$ and $\sigma$ are the mean and standard deviation of the QLD, respectively. We present a full derivation of this analytical solution in Appendix~\ref{sec:app}, to which we refer the interested reader.

\begin{figure}
\centering
\includegraphics[width=\columnwidth]{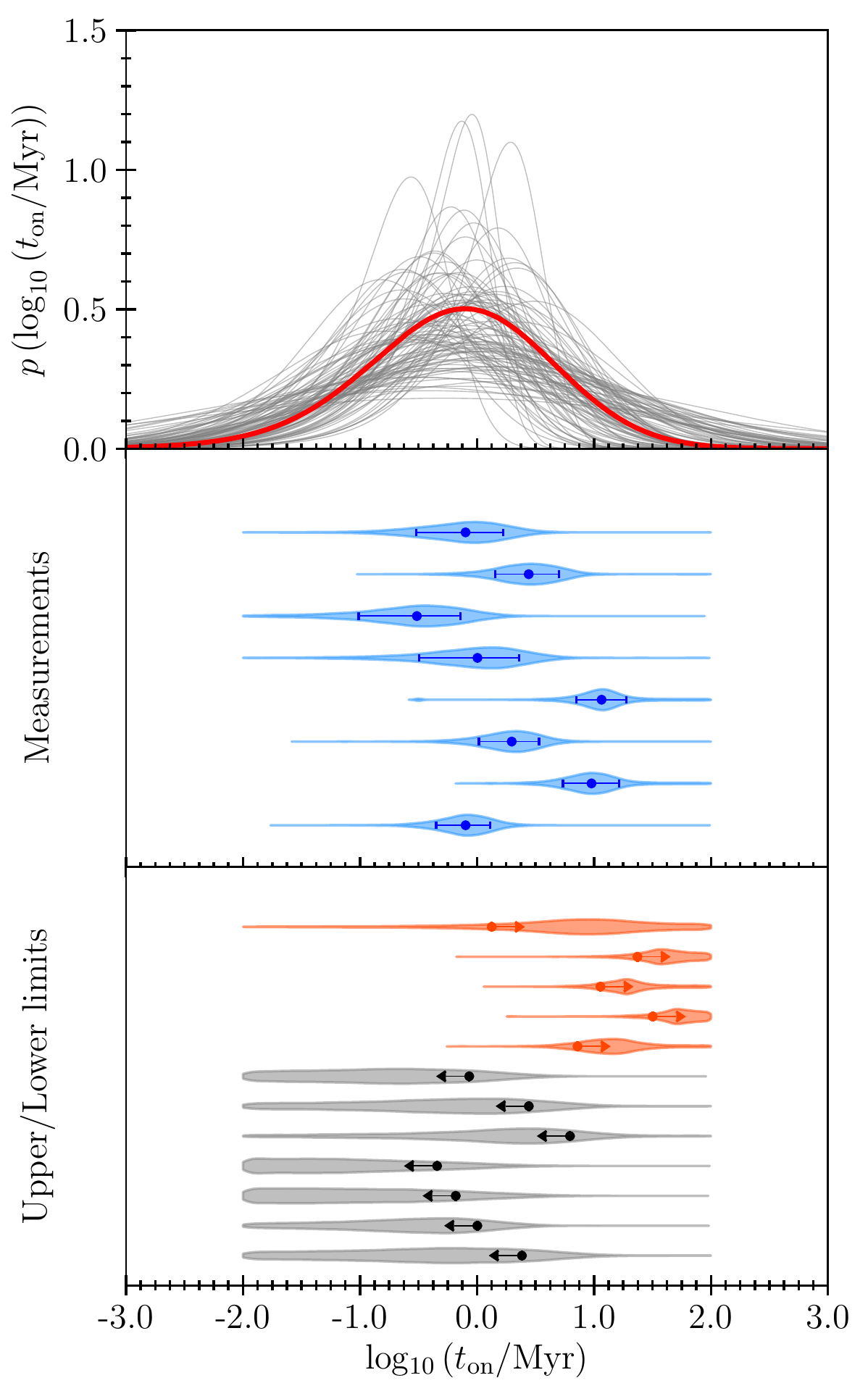}
\caption{\label{fig:violin}
 Distribution of quasar on-times $p\left( {\rm log_{10}}\left(t_{\rm on}/{\rm Myr}\right)\right)$. The grey curves in the top panel show $100$ realizations of the $p\left( {\rm log_{10}}\left(t_{\rm on}/{\rm Myr}\right)\right)$ calculated using eq.~(\ref{eq:ansol}), based on the values of QLD mean $\mu$ and standard deviation $\sigma$ illustrated in the right panel of Fig.~\ref{fig:mcmc}. The solid red curve illustrates the $p\left( {\rm log_{10}}\left(t_{\rm on}/{\rm Myr}\right)\right)$ distribution corresponding to the maximum likelihood values of $\mu$ and $\sigma$. The middle panel shows the measured on-times of individual quasars, while the bottom panel illustrates the upper/lower limits on $ {\rm log_{10}}\left(t_{\rm on}/{\rm Myr}\right)$ (see Table~\ref{tab:table1}). The violins show the corresponding posterior distributions inferred from the analysis of individual quasars (see \citetalias{Khrykin2019} and \citetalias{Worseck2021} for details).}
\end{figure}

The top panel of Fig.~\ref{fig:violin} illustrates the resulting PDFs computed using eq.~(\ref{eq:ansol}). The grey lines show $100$ realizations of the $p\left( {\rm log}_{10}\left( t_{\rm on} / {\rm Myr} \right) | \mu , \sigma \right)$ for $100$ pairs of the $\{ \mu , \sigma \}$ values randomly drawn from the MCMC chains (see Section~\ref{sec:mcmc} for details). The solid red line displays the quasar on-time probability distribution function corresponding to the maximum likelihood values of the QLD parameters (see red curve in the right panel of Fig.~\ref{fig:mcmc}), and calculated using eq.~(\ref{eq:ansol}). We compare these results to the on-time constraints of individual quasars (see Table~\ref{tab:table1}) in the middle and bottom panels of Fig.~\ref{fig:violin}.

\begin{figure}
\centering
\includegraphics[width=\columnwidth]{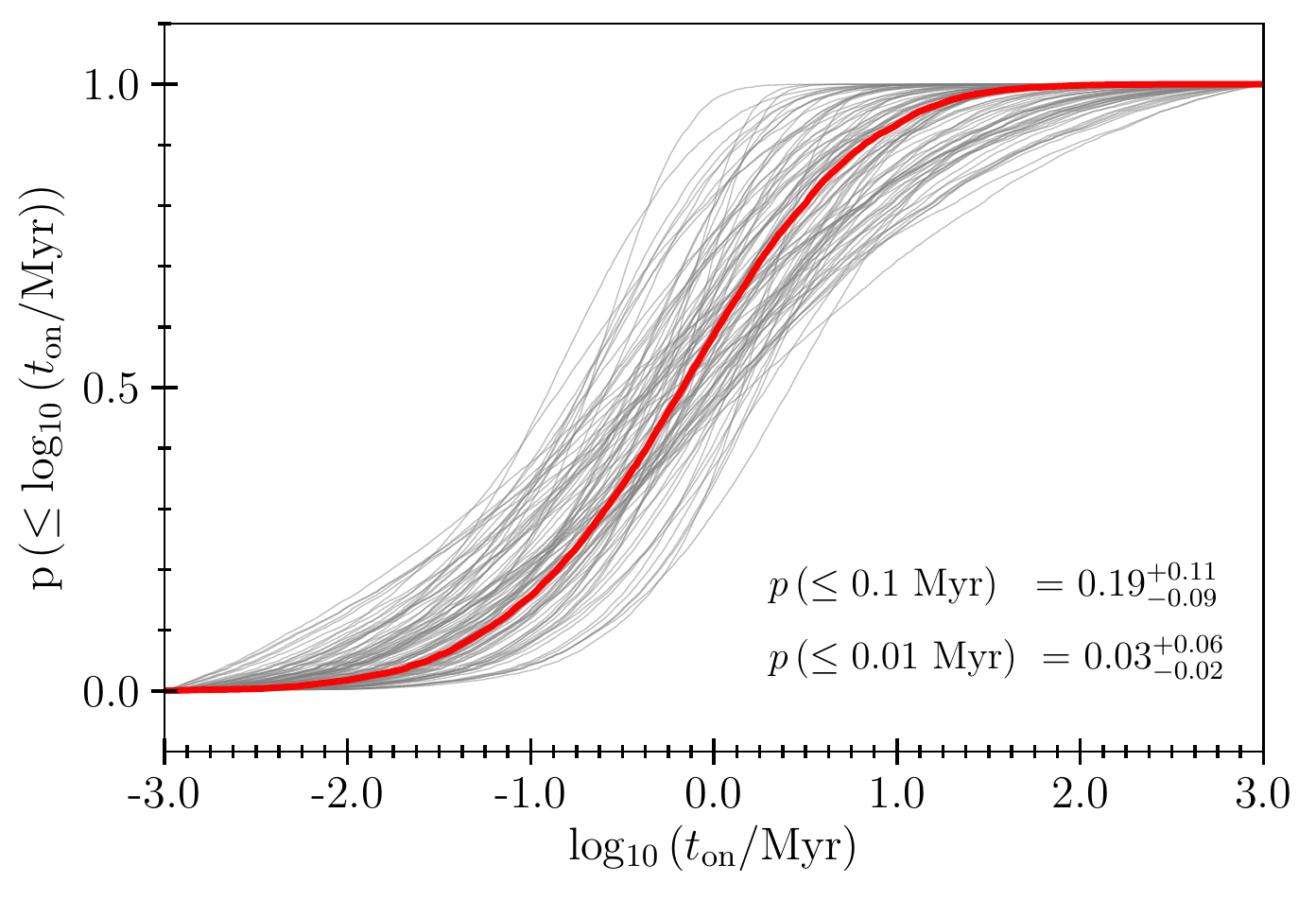}
\caption{\label{fig:ton_cdf}
The CDF of the quasar on-times. The grey curves are the CDF realizations corresponding to $100$ realizations of the $p\left( {\rm log_{10}}\left(t_{\rm on}/{\rm Myr}\right)\right)$, illustrated in Fig.~\ref{fig:violin}, while the red curve is the CDF corresponding to the maximum likelihood values of the QLD parameters. We also quote the estimated fractions of quasars with on-time shorter than $0.1$ and $0.01$~Myr.}
\end{figure}

It is apparent from the top panel of Fig.~\ref{fig:violin} that our analytical solution for the PDF of the quasar on-times has a tail towards low values of $t_{\rm on}$, which results from the uniform sampling of $t_{\rm on}$ given $t_{\rm Q}$, i.e., $t_{\rm on} \sim \mathcal{U}\left[ 0, t_{\rm Q} \right]$. The existence of this tail may explain the discovery of very young quasars at $z_{\rm qso} \simeq 6$, assuming that quasars at higher redshift follow the same distribution. Indeed, \citet{Eilers2020} analysed the \ion{H}{i} Ly$\alpha$ proximity zones in the spectra of $153$ quasars at $z_{\rm qso} \simeq 6$ and found that the probability of finding quasars with $t_{\rm on} \leq 0.1$~Myr is $p \left( \leq 0.1~{\rm Myr}\right) \simeq 0.05$--$0.10$. We can estimate the young quasar fraction from the inferred distributions of $t_{\rm on}$ times by integrating 

\begin{equation}
\label{eq:cdf}
    p\left(\leq 0.1~{\rm Myr} \right) = \int_{-\infty}^{-1} p\left( {\rm log}_{10}\left( t_{\rm on} / {\rm Myr} \right) \right) {\rm d}~{\rm log_{10}}\left(t_{\rm on}/{\rm Myr}\right),
\end{equation}
where $p\left( {\rm log}_{10}\left( t_{\rm on} / {\rm Myr} \right) \right)$ is given by eq.~(\ref{eq:ansol}). Fig.~\ref{fig:ton_cdf} shows the resulting cumulative probability distributions (CDF) calculated from the same realizations of the $t_{\rm on}$ PDF illustrated in the top panel of Fig.~\ref{fig:violin}. Applying eq.~(\ref{eq:cdf}) to $\sim 13700$ realizations of the quasar $t_{\rm on}$ PDFs (based on total number of $\{ \mu , \sigma \}$ pairs from MCMC samples) we infer $p \left( \leq 0.1~{\rm Myr}\right) = 0.19^{+0.11}_{-0.09}$. This is a factor of $2$--3 higher than the value estimated by \citet{Eilers2020} who found $p \left( \leq 0.1~{\rm Myr}\right) = 0.05-0.10$. However, we emphasize that the value quoted by \citet{Eilers2020} is likely a lower limit due to possible incompleteness of their sample, and thus our estimate is broadly consistent. Future analysis of a larger sample of observed \ion{He}{ii} proximity zones should help to refine this estimate. In addition, the properties of the underlying QLD might vary with redshift, which will also affect the estimated fraction of young quasars.

Finally, we note that the short quasar on-times and quasar lifetimes implied by our inferred QLD ($ \log_{10}\left( {t_{\rm Q}} / {\rm Myr} \right) = 0.22^{+0.22}_{-0.25}$ and $\sigma_{\log_{10}t_{\rm Q}} = 0.80^{+0.37}_{-0.27}$; see Section~\ref{sec:mcmc}) appear to be at odds with measurements obtained from the analysis of the quasar clustering. Indeed, \citet{Shen2007} estimated the fraction of the age of the Universe that a galaxy hosts an active quasar, i.e., the duty cycle $f_{\rm dc}$, at $z \geq 3.5$ to be $f_{\rm dc} \simeq 3-60\%$ (see also \citealp{White2008}, \citealp{Shankar2010}). This measurement translates into the broad range of integrated quasar lifetimes $t_{\rm dc} \simeq f_{\rm dc}~t_{\rm H}(z) \simeq 30-600$~Myr, where $t_{\rm H}(z)$ is the age of the Universe at redshift $z$. These time-scales are $\approx 10-200$ times longer than the lifetimes suggested by our results \citep[see also][]{He2018}. If quasars are to radiate their energy in only one continuous "light-bulb" burst lasting for $t_{\rm Q} = t_{\rm dc} \simeq f_{\rm dc}~t_{\rm H}\left( z \right)$, then there is no simple way to resolve the apparent discrepancy between our results and clustering measurements. However, it needs to be noted that the relationship between the estimated clustering strength and the inferred quasar lifetimes is a subject of significant uncertainty due to our lack of knowledge of how quasars populate the dark matter haloes \citep{Shen2007, White2008, He2018}.

\subsection{Implications for SMBH growth}
\label{sec:qld_smbh}

The short quasar lifetimes that our findings point to pose a significant challenge to the models of SMBH formation and evolution. Indeed, in the simplest model of constant accretion at the Eddington limit, the SMBH mass follows an exponential growth law given by
\begin{equation}
\label{eq:smbh}
    M_{\rm SMBH} = M_{\rm seed} \cdot {\rm exp}\left( \frac{t-t_{\rm seed}}{t_{\rm S}} \right),
\end{equation}
where $M_{\rm seed}$ is the initial mass of the SMBH at time $t_{\rm seed}$, $t_{\rm S} \simeq 450 \cdot {\rm \lambda}^{-1} \epsilon \slash \left( 1 - \epsilon \right) $~Myr is the $e$-folding or Salpeter time-scale, ${\rm \lambda} = L_{\rm bol} \slash L_{\rm edd}$ is the Eddington ratio, and $\epsilon$ is the radiative efficiency of accretion which we set to the canonical value from general relativity \citep{Thorne1974} of $\epsilon=0.1$ \citep[but see][]{Davies2019}. 

Let us assume that SMBHs in quasars grow only via emitting UV radiation, i.e., there are no significant periods of obscured black hole growth. Assuming an Eddington ratio of unity, a $10$~per cent radiative efficiency and a stellar-mass SMBH seed of $M_{\rm seed}\simeq 100\ M_{\sun}$, then, according to eq.~(\ref{eq:smbh}), SMBH must constantly accrete for a total of $\simeq 16$ Salpeter $e$-folding time-scales or $\simeq 730$~Myr in order to grow the $M_{\rm SMBH} \simeq 10^9\ {\rm M}_{\sun}$ SMBH masses measured for our sample of quasars \citepalias{Worseck2021}. It is well known that this poses a problem for quasars at $z\simeq 6-7$ where the $< 1$~Gyr age of the Universe implies there is too little time to grow the observed $\gtrsim 10^9 M_\odot$ SMBHs \citep{Mortlock2011, Banados2018, Wang2020, Yang2020a, Wang2021} from the stellar-mass seeds. 

At fact value, the $\sim 730~{\rm Myr}$ required to grow a $\sim 10^9\,M_\odot$ appears inconsistent with both the measured short on-times of individual quasars and the inferred QLD, which also favors much shorter quasar lifetimes (see Fig.~\ref{fig:mcmc}). However, we note that \ion{He}{ii} proximity zones are not sensitive to this entire growth history of $\simeq 16$ Salpeter times given that the SMBH is much smaller and the quasar is much fainter in the distant past.
Only the last e-folding time has an influence on the extent of the \ion{He}{ii} proximity zone. As such, the most fair comparison of our assumed "light-bulb" light curve to this exponential light curve is to approximately commensurate $t_{\rm Q}$ with the $e$-folding time-scale $t_{\rm S}$, i.e., by simply equating the total area under these respective curves, that is $\int L \left( t \right) {\rm d}t = t_{\rm S}~L_0 \approx t_{\rm Q}~L_0$, where $L_0$ is the luminosity of the quasar at the time (redshift) at which we observe the quasar. This would seem to imply the Salpeter times of $t_S \simeq 1$~Myr or about $50$ times shorter than the canonical value, thus requiring a commensurate reduction of the radiative efficiency to $\epsilon \simeq 0.002$. \citet{Davies2019} made a similar argument to explain the short quasar lifetimes implied by $z \simeq 7$ proximity zones. Such low radiative efficiencies would mean that SMBH can grow extremely rapidly, i.e., $16$ Salpeter times would correspond to $16$~Myr. If this also holds at high-z as suggested by \citet{Davies2019}, then there would be no problem growing $M_{\rm SMBH} \simeq 10^9~M_{\sun}$ in less than 1~Gyr after the Big Bang. 

However, application of the Soltan argument \citep{Soltan1982} that relates the total UV emission from quasars to the local SMBH population, seems to be consistent with $\epsilon = 0.1$ \citep{Yu2002, Shankar2009, Ueda2014}. Thus, the much smaller radiative efficiencies suggested by $z\sim 3$ quasar \ion{He}{ii} proximity zone sizes would then result in the over-production of SMBHs relative to what is observed locally. For \citet{Davies2019} this was not as big of a problem since $z \simeq 7$ quasars do not contribute significantly to the total quasar luminosity density that informs the Soltan argument, and one can simply invoke lower radiative efficiencies for these extreme early SMBHs. This is not the case for our sample of representative quasars at $z \simeq 3-4$, and thus our results appear in conflict with the Soltan argument -- the short $e$-folding time-scale and corresponding low radiative efficiency required by proximity zones would  seem to imply that there exists a large mass density of unseen SMBHs in the local universe. There are two potential mechanisms to  resolve this apparent conflict, namely 
one can invoke significant periods of UV-obscured growth or more complex light curves, which we now discuss in turn. 

With regards to obscuration, suppose that the $\approx 1$~Myr light-bulb quasar lifetimes that we measure can be thought of as an ephemeral UV-unobscured phase of a predominantly obscured black hole growth history. Assuming the standard model with $\epsilon = 0.1$ and $t_{\rm S} = 45$~Myr, the obscured to unobscured ratio at comparable luminosity would then need to be $t_{\rm S}/t_{\rm Q}$. This would thus imply about $50$ obscured quasars to every unobscured one. However, observations at $z \simeq 3$ have not uncovered such large number of obscured quasars. Indeed, \citet{Poletta2008} and \citet{Merloni2014} find a roughly comparable number of obscured to unobscured quasars at $z\sim 2-3$ (but see \citealp{Vito2018}). It seems unlikely that an obscured population $\simeq 50$ times larger could have gone unnoticed, implying that predominantly obscured black hole growth does not obviously resolve the aforementioned tension with the Soltan argument.

On the other hand, SMBH growth might not obey the simple continuous Eddington-limited exponential evolution described by eq.~(\ref{eq:smbh}). Indeed, more complex flickering light curves, characterized by short time-scale variations in the quasar continuum radiation \citep{Ciotti2001, Novak2011}, could manage to grow the observed SMBHs and remain consistent with our proximity zone constraints. For the sake of illustration, let us assume a simple flickering light curve whereby quasars turn on and shine for a constant periods equal to the light bulb $t_{\rm Q}$ drawn from the QLD, but then turn off for extended periods denoted by $t_{\rm off}$. Even for this simple on-off light curve the space of possibilities is large. We, however, focus on the simplest variant, which is that the quasars shine as light bulbs for $t_{\rm Q}$ and then are off for $t_{\rm off} = t_{\rm eq}$, and that this behavior continues for the entire Hubble time $t_{\rm H}$ ($\simeq 1.6$~Gyr at $z\simeq 4$). This would result in the \ion{He}{ii} proximity zone size distributions that are effectively identical to those simulated in this work, because the long $t_{\rm off}$ allows the proximity zone sizes to decay to zero, which occurs on the equilibration time-scale $t_{\rm eq}$ as shown by \citet{Davies2020a} for analogous \ion{H}{i} proximity zones. In this picture the effective duty cycle for UV-luminous phases would be $t_{\rm dc} = t_{\rm Q}/t_{\rm off}\cdot t_{\rm H}$, which for $t_{\rm Q} = 1-10$~Myr and $t_{\rm off} = t_{\rm eq} \approx 30$~Myr implies $t_{\rm dc} \simeq 50-500$~Myr. Are duty cycles in this range inconsistent with clustering constraints on the duty cycle and SMBH growth? For the former, we note that these duty cycles overlap the range implied by quasar clustering, which allow for duty cycles in the range $30-600$~Myr \citep{Shen2007, White2008, Shankar2010, He2018}. Regarding SMBH growth, it appears that this would not be enough time to growth the SMBHs given the standard model of $t_{\rm S} = 45$~Myr, since roughly $16$ Salpeter times are required, or about $730$~Myr. 

There thus appears to be a tension between the inferred quasar lifetimes and the SMBH evolution models, although we emphasize that: 1) we have assumed that this on/off behavior also applies to early phases of SMBH growth when the quasars are much fainter,
which our proximity zone observations do not constrain; 2) we have considered a single value of $t_{\rm on}/t_{\rm off}$ whereas the space of possible combinations is large, and light curves could surely be significantly more complex; 3) we have not considered the effects of obscured growth phases, which current constraints on obscured populations suggest could modify the light curve math at the factor of a few level (but not significantly more). 

Further careful work on this question is clearly warranted, and we conclude by noting that it is quite possible that a similar SMBH growth problem that has been touted at $z\gtrsim 7$ exists even at $z\sim 3$. Whereas for the standard model of exponential growth one would conclude that $16\cdot t_{\rm S} \approx 730$~Myr is required to grow SMBHs, which seems straightforward given the $\approx 1.6$~Gyr age of the Universe available, the constraints from the proximity zone sizes indicate that you cannot emit UV radiation over this entire time period lest the \ion{He}{ii} proximity zones appear much larger. Hence, there is considerably less time available to grow the SMBHs than the full age of the Universe at $z\sim 3$. 

\section{Conclusions}
\label{sec:conc}

We have used the measured \ion{He}{ii} proximity zone sizes of $N_{\rm qso} = 20$ quasars at $2.7 \lesssim z_{\rm qso} \lesssim 3.9$ \citepalias[][\citetalias{Worseck2021}]{Khrykin2019} to infer the intrinsic quasar lifetime distribution. We have created a new fully Bayesian MCMC formalism that performs statistical comparison between the sizes of observed \ion{He}{ii} proximity zones to the outcome of the radiative transfer simulations. This allows us to infer the shape of the underlying distribution of quasar lifetimes $t_{\rm Q}$ for the first time, marginalized over the ionization state of \ion{He}{ii} in the surrounding intergalactic medium. The main results of our work are as follows:

\begin{enumerate}

    \item Assuming a log-normal distribution of quasar lifetimes, we inferred the mean of the distribution  $\langle {\rm log}_{10} \left( t_{\rm Q} / {\rm Myr} \right) \rangle = 0.22^{+0.22}_{-0.25}$, and the standard deviation $\sigma_{{\rm log}_{10}t_{\rm Q }} = 0.80^{+0.37}_{-0.27}$ (see Figure~\ref{fig:mcmc}).
    
    \item We presented a simple analytical expression for the distribution of the quasar on-times $t_{\rm on}$. We showed that according to this distribution, the probability of finding quasars with $t_{\rm on} \leq 0.1$~Myr is $p \left( \leq 0.1~{\rm Myr}\right) = 0.19^{+0.11}_{-0.09}$, broadly consistent with the values found by \citet{Eilers2020} based on the analysis of the \ion{H}{i} proximity zones at $z_{\rm qso} \simeq 6$.
    
    \item We also analysed the impact of redshift uncertainties on the results of our inference algorithm. We found that in contrast to individual measurements of quasar on-times (see \citetalias{Khrykin2019} and \citetalias{Worseck2021}), these uncertainties do not  significantly modify the constrained parameters of the quasar lifetime distribution.

    \item We discussed our inferred quasar lifetime distribution in the context of SMBH growth and found apparent tension between our estimates and the standard model of exponential growth. We noted that several possible solutions to this tension, including invoking periods of UV-obscured growth and flickering light-curves, do not necessarily solve this tension. Therefore, a more careful modelling is required in the future to resolve this problem.
    
\end{enumerate}

We note that the precision with which we can constrain the quasar lifetime distribution depends on the number of quasars in the analysed sample. Therefore, future progress requires more high-quality data. Our ongoing HST observational campaign (Program 16318, PI: Worseck) aims to double the sample of quasars with measured \ion{He}{ii} proximity zones at $z_\mathrm{qso}\simeq 2.7$--$3.2$, significantly improving the precision of our inference algorithm. Finally, we emphasize that there also exists a large sample ($\simeq 150$) of measured \ion{H}{i} proximity zones at $z_{\rm qso} \simeq 6$ \citep{Eilers2020, Morey2021}. The same inference algorithm presented in this work can be applied to this higher redshift sample. Besides improved statistics, such analysis opens up the possibility of constraining the redshift evolution of the quasar lifetime distribution parameters.

\section*{Acknowledgements}

IKS thanks Metin Ata, Valeri Vardanyan, and Anna-Christina Eilers for useful discussions. Kavli IPMU is supported by World Premier International Research Center Initiative (WPI), MEXT, Japan. JFH acknowledges support for this work that was provided by NASA through grant numbers HST-AR-15014.003-A and HST-GO-16318.002-A from the Space Telescope Science Institute, which is operated by AURA, Inc., under NASA contract NAS 5-26555.

\section*{Data availability}

The data underlying this article will be shared on reasonable request to the corresponding author.




\typeout{}
\bibliographystyle{mnras}
\bibliography{references} 

\begin{thebibliography}{}
\makeatletter
\relax
\def\mn@urlcharsother{\let\do\@makeother \do\$\do\&\do\#\do\^\do\_\do\%\do\~}
\def\mn@doi{\begingroup\mn@urlcharsother \@ifnextchar [ {\mn@doi@}
  {\mn@doi@[]}}
\def\mn@doi@[#1]#2{\def\@tempa{#1}\ifx\@tempa\@empty \href
  {http://dx.doi.org/#2} {doi:#2}\else \href {http://dx.doi.org/#2} {#1}\fi
  \endgroup}
\def\mn@eprint#1#2{\mn@eprint@#1:#2::\@nil}
\def\mn@eprint@arXiv#1{\href {http://arxiv.org/abs/#1} {{\tt arXiv:#1}}}
\def\mn@eprint@dblp#1{\href {http://dblp.uni-trier.de/rec/bibtex/#1.xml}
  {dblp:#1}}
\def\mn@eprint@#1:#2:#3:#4\@nil{\def\@tempa {#1}\def\@tempb {#2}\def\@tempc
  {#3}\ifx \@tempc \@empty \let \@tempc \@tempb \let \@tempb \@tempa \fi \ifx
  \@tempb \@empty \def\@tempb {arXiv}\fi \@ifundefined
  {mn@eprint@\@tempb}{\@tempb:\@tempc}{\expandafter \expandafter \csname
  mn@eprint@\@tempb\endcsname \expandafter{\@tempc}}}

\bibitem[\protect\citeauthoryear{{Anderson}, {Hogan}, {Williams}  \&
  {Carswell}}{{Anderson} et~al.}{1999}]{Anderson1999}
{Anderson} S.~F.,  {Hogan} C.~J.,  {Williams} B.~F.,   {Carswell} R.~F.,  1999,
  \mn@doi [\aj] {10.1086/300698}, \href
  {https://ui.adsabs.harvard.edu/abs/1999AJ....117...56A} {117, 56}

\bibitem[\protect\citeauthoryear{{Angl{\'e}s-Alc{\'a}zar}, {{\"O}zel}  \&
  {Dav{\'e}}}{{Angl{\'e}s-Alc{\'a}zar} et~al.}{2013}]{Angles-alcazar2013}
{Angl{\'e}s-Alc{\'a}zar} D.,  {{\"O}zel} F.,   {Dav{\'e}} R.,  2013, \mn@doi
  [\apj] {10.1088/0004-637X/770/1/5}, \href
  {https://ui.adsabs.harvard.edu/abs/2013ApJ...770....5A} {770, 5}

\bibitem[\protect\citeauthoryear{{Angl{\'e}s-Alc{\'a}zar},
  {Faucher-Gigu{\`e}re}, {Quataert}, {Hopkins}, {Feldmann}, {Torrey}, {Wetzel}
  \& {Kere{\v{s}}}}{{Angl{\'e}s-Alc{\'a}zar} et~al.}{2017}]{Angles-alcazar2017}
{Angl{\'e}s-Alc{\'a}zar} D.,  {Faucher-Gigu{\`e}re} C.-A.,  {Quataert} E.,
  {Hopkins} P.~F.,  {Feldmann} R.,  {Torrey} P.,  {Wetzel} A.,   {Kere{\v{s}}}
  D.,  2017, \mn@doi [\mnras] {10.1093/mnrasl/slx161}, \href
  {https://ui.adsabs.harvard.edu/abs/2017MNRAS.472L.109A} {472, L109}

\bibitem[\protect\citeauthoryear{{Angles-Alcazar} et~al.,}{{Angles-Alcazar}
  et~al.}{2020}]{Angles2020}
{Angles-Alcazar} D.,  et~al., 2020, arXiv e-prints, \href
  {https://ui.adsabs.harvard.edu/abs/2020arXiv200812303A} {p. arXiv:2008.12303}

\bibitem[\protect\citeauthoryear{{Ba{\~n}ados} et~al.,}{{Ba{\~n}ados}
  et~al.}{2018}]{Banados2018}
{Ba{\~n}ados} E.,  et~al., 2018, \mn@doi [\nat] {10.1038/nature25180}, \href
  {https://ui.adsabs.harvard.edu/abs/2018Natur.553..473B} {553, 473}

\bibitem[\protect\citeauthoryear{{Ba{\~n}ados} et~al.,}{{Ba{\~n}ados}
  et~al.}{2019}]{Banados2019}
{Ba{\~n}ados} E.,  et~al., 2019, \mn@doi [\apjl] {10.3847/2041-8213/ab3659},
  \href {https://ui.adsabs.harvard.edu/abs/2019ApJ...881L..23B} {881, L23}

\bibitem[\protect\citeauthoryear{{Bajtlik}, {Duncan}  \& {Ostriker}}{{Bajtlik}
  et~al.}{1988}]{Bajtlik1988}
{Bajtlik} S.,  {Duncan} R.~C.,   {Ostriker} J.~P.,  1988, \mn@doi [\apj]
  {10.1086/166217}, \href
  {https://ui.adsabs.harvard.edu/abs/1988ApJ...327..570B} {327, 570}

\bibitem[\protect\citeauthoryear{{Baldwin}}{{Baldwin}}{1977}]{Baldwin1977}
{Baldwin} J.~A.,  1977, \mn@doi [\apj] {10.1086/155294}, \href
  {https://ui.adsabs.harvard.edu/abs/1977ApJ...214..679B} {214, 679}

\bibitem[\protect\citeauthoryear{{Becker} \& {Bolton}}{{Becker} \&
  {Bolton}}{2013}]{Becker2013}
{Becker} G.~D.,  {Bolton} J.~S.,  2013, \mn@doi [\mnras]
  {10.1093/mnras/stt1610}, \href
  {https://ui.adsabs.harvard.edu/abs/2013MNRAS.436.1023B} {436, 1023}

\bibitem[\protect\citeauthoryear{{Bosman}, {Fan}, {Jiang}, {Reed}, {Matsuoka},
  {Becker}  \& {Haehnelt}}{{Bosman} et~al.}{2018}]{Bosman2018}
{Bosman} S. E.~I.,  {Fan} X.,  {Jiang} L.,  {Reed} S.,  {Matsuoka} Y.,
  {Becker} G.,   {Haehnelt} M.,  2018, \mn@doi [\mnras]
  {10.1093/mnras/sty1344}, \href
  {https://ui.adsabs.harvard.edu/abs/2018MNRAS.479.1055B} {479, 1055}

\bibitem[\protect\citeauthoryear{{Bournaud}, {Dekel}, {Teyssier}, {Cacciato},
  {Daddi}, {Juneau}  \& {Shankar}}{{Bournaud} et~al.}{2011}]{Bournaud2011}
{Bournaud} F.,  {Dekel} A.,  {Teyssier} R.,  {Cacciato} M.,  {Daddi} E.,
  {Juneau} S.,   {Shankar} F.,  2011, \mn@doi [\apjl]
  {10.1088/2041-8205/741/2/L33}, \href
  {https://ui.adsabs.harvard.edu/abs/2011ApJ...741L..33B} {741, L33}

\bibitem[\protect\citeauthoryear{{Capelo}, {Volonteri}, {Dotti}, {Bellovary},
  {Mayer}  \& {Governato}}{{Capelo} et~al.}{2015}]{Capelo2015}
{Capelo} P.~R.,  {Volonteri} M.,  {Dotti} M.,  {Bellovary} J.~M.,  {Mayer} L.,
   {Governato} F.,  2015, \mn@doi [\mnras] {10.1093/mnras/stu2500}, \href
  {https://ui.adsabs.harvard.edu/abs/2015MNRAS.447.2123C} {447, 2123}

\bibitem[\protect\citeauthoryear{{Carswell}, {Whelan}, {Smith}, {Boksenberg}
  \& {Tytler}}{{Carswell} et~al.}{1982}]{Carswell1982}
{Carswell} R.~F.,  {Whelan} J.~A.~J.,  {Smith} M.~G.,  {Boksenberg} A.,
  {Tytler} D.,  1982, \mn@doi [\mnras] {10.1093/mnras/198.1.91}, \href
  {https://ui.adsabs.harvard.edu/abs/1982MNRAS.198...91C} {198, 91}

\bibitem[\protect\citeauthoryear{{Cen} \& {Safarzadeh}}{{Cen} \&
  {Safarzadeh}}{2015}]{Cen2015}
{Cen} R.,  {Safarzadeh} M.,  2015, \mn@doi [\apjl]
  {10.1088/2041-8205/798/2/L38}, \href
  {https://ui.adsabs.harvard.edu/abs/2015ApJ...798L..38C} {798, L38}

\bibitem[\protect\citeauthoryear{{Chardin}, {Puchwein}  \&
  {Haehnelt}}{{Chardin} et~al.}{2017}]{Chardin2017}
{Chardin} J.,  {Puchwein} E.,   {Haehnelt} M.~G.,  2017, \mn@doi [\mnras]
  {10.1093/mnras/stw2943}, \href
  {https://ui.adsabs.harvard.edu/abs/2017MNRAS.465.3429C} {465, 3429}

\bibitem[\protect\citeauthoryear{{Choi}, {Somerville}, {Ostriker}, {Naab}  \&
  {Hirschmann}}{{Choi} et~al.}{2018}]{Choi2018}
{Choi} E.,  {Somerville} R.~S.,  {Ostriker} J.~P.,  {Naab} T.,   {Hirschmann}
  M.,  2018, \mn@doi [\apj] {10.3847/1538-4357/aae076}, \href
  {https://ui.adsabs.harvard.edu/abs/2018ApJ...866...91C} {866, 91}

\bibitem[\protect\citeauthoryear{{Ciotti} \& {Ostriker}}{{Ciotti} \&
  {Ostriker}}{2001}]{Ciotti2001}
{Ciotti} L.,  {Ostriker} J.~P.,  2001, \mn@doi [\apj] {10.1086/320053}, \href
  {https://ui.adsabs.harvard.edu/abs/2001ApJ...551..131C} {551, 131}

\bibitem[\protect\citeauthoryear{{Compostella}, {Cantalupo}  \&
  {Porciani}}{{Compostella} et~al.}{2013}]{Compostella2013}
{Compostella} M.,  {Cantalupo} S.,   {Porciani} C.,  2013, \mn@doi [\mnras]
  {10.1093/mnras/stt1510}, \href
  {https://ui.adsabs.harvard.edu/abs/2013MNRAS.435.3169C} {435, 3169}

\bibitem[\protect\citeauthoryear{{Conroy} \& {White}}{{Conroy} \&
  {White}}{2013}]{Conroy2013}
{Conroy} C.,  {White} M.,  2013, \mn@doi [\apj] {10.1088/0004-637X/762/2/70},
  \href {https://ui.adsabs.harvard.edu/abs/2013ApJ...762...70C} {762, 70}

\bibitem[\protect\citeauthoryear{{D'Aloisio}, {Upton Sanderbeck}, {McQuinn},
  {Trac}  \& {Shapiro}}{{D'Aloisio} et~al.}{2017}]{DAloisio2017}
{D'Aloisio} A.,  {Upton Sanderbeck} P.~R.,  {McQuinn} M.,  {Trac} H.,
  {Shapiro} P.~R.,  2017, \mn@doi [\mnras] {10.1093/mnras/stx711}, \href
  {https://ui.adsabs.harvard.edu/abs/2017MNRAS.468.4691D} {468, 4691}

\bibitem[\protect\citeauthoryear{{Dall'Aglio}, {Wisotzki}  \&
  {Worseck}}{{Dall'Aglio} et~al.}{2008}]{Dallaglio2008}
{Dall'Aglio} A.,  {Wisotzki} L.,   {Worseck} G.,  2008, \mn@doi [\aap]
  {10.1051/0004-6361:200810724}, \href
  {https://ui.adsabs.harvard.edu/abs/2008A&A...491..465D} {491, 465}

\bibitem[\protect\citeauthoryear{{Davies}, {Furlanetto}  \& {Dixon}}{{Davies}
  et~al.}{2017}]{Davies2017}
{Davies} F.~B.,  {Furlanetto} S.~R.,   {Dixon} K.~L.,  2017, \mn@doi [\mnras]
  {10.1093/mnras/stw2868}, \href
  {https://ui.adsabs.harvard.edu/abs/2017MNRAS.465.2886D} {465, 2886}

\bibitem[\protect\citeauthoryear{{Davies}, {Hennawi}  \& {Eilers}}{{Davies}
  et~al.}{2019}]{Davies2019}
{Davies} F.~B.,  {Hennawi} J.~F.,   {Eilers} A.-C.,  2019, \mn@doi [\apjl]
  {10.3847/2041-8213/ab42e3}, \href
  {https://ui.adsabs.harvard.edu/abs/2019ApJ...884L..19D} {884, L19}

\bibitem[\protect\citeauthoryear{{Davies}, {Hennawi}  \& {Eilers}}{{Davies}
  et~al.}{2020}]{Davies2020a}
{Davies} F.~B.,  {Hennawi} J.~F.,   {Eilers} A.-C.,  2020, \mn@doi [\mnras]
  {10.1093/mnras/stz3303}, \href
  {https://ui.adsabs.harvard.edu/abs/2020MNRAS.493.1330D} {493, 1330}

\bibitem[\protect\citeauthoryear{{Dayal} et~al.,}{{Dayal}
  et~al.}{2020}]{Dayal2020}
{Dayal} P.,  et~al., 2020, \mn@doi [\mnras] {10.1093/mnras/staa1138}, \href
  {https://ui.adsabs.harvard.edu/abs/2020MNRAS.495.3065D} {495, 3065}

\bibitem[\protect\citeauthoryear{{De Rosa} et~al.,}{{De Rosa}
  et~al.}{2014}]{DeRosa2014}
{De Rosa} G.,  et~al., 2014, \mn@doi [\apj] {10.1088/0004-637X/790/2/145},
  \href {https://ui.adsabs.harvard.edu/abs/2014ApJ...790..145D} {790, 145}

\bibitem[\protect\citeauthoryear{{Di Matteo}, {Springel}  \& {Hernquist}}{{Di
  Matteo} et~al.}{2005}]{DiMatteo2005}
{Di Matteo} T.,  {Springel} V.,   {Hernquist} L.,  2005, \mn@doi [\nat]
  {10.1038/nature03335}, \href
  {https://ui.adsabs.harvard.edu/abs/2005Natur.433..604D} {433, 604}

\bibitem[\protect\citeauthoryear{{Eftekharzadeh} et~al.,}{{Eftekharzadeh}
  et~al.}{2015}]{Eftekharzadeh2015}
{Eftekharzadeh} S.,  et~al., 2015, \mn@doi [\mnras] {10.1093/mnras/stv1763},
  \href {https://ui.adsabs.harvard.edu/abs/2015MNRAS.453.2779E} {453, 2779}

\bibitem[\protect\citeauthoryear{{Eilers}, {Davies}, {Hennawi}, {Prochaska},
  {Luki{\'c}}  \& {Mazzucchelli}}{{Eilers} et~al.}{2017}]{Eilers2017}
{Eilers} A.-C.,  {Davies} F.~B.,  {Hennawi} J.~F.,  {Prochaska} J.~X.,
  {Luki{\'c}} Z.,   {Mazzucchelli} C.,  2017, \mn@doi [\apj]
  {10.3847/1538-4357/aa6c60}, \href
  {https://ui.adsabs.harvard.edu/abs/2017ApJ...840...24E} {840, 24}

\bibitem[\protect\citeauthoryear{{Eilers}, {Hennawi}  \& {Davies}}{{Eilers}
  et~al.}{2018}]{Eilers2018}
{Eilers} A.-C.,  {Hennawi} J.~F.,   {Davies} F.~B.,  2018, \mn@doi [\apj]
  {10.3847/1538-4357/aae081}, \href
  {https://ui.adsabs.harvard.edu/abs/2018ApJ...867...30E} {867, 30}

\bibitem[\protect\citeauthoryear{{Eilers} et~al.,}{{Eilers}
  et~al.}{2020}]{Eilers2020}
{Eilers} A.-C.,  et~al., 2020, \mn@doi [\apj] {10.3847/1538-4357/aba52e}, \href
  {https://ui.adsabs.harvard.edu/abs/2020ApJ...900...37E} {900, 37}

\bibitem[\protect\citeauthoryear{{Fan} et~al.,}{{Fan} et~al.}{2006}]{Fan2006}
{Fan} X.,  et~al., 2006, \mn@doi [\aj] {10.1086/504836}, \href
  {https://ui.adsabs.harvard.edu/abs/2006AJ....132..117F} {132, 117}

\bibitem[\protect\citeauthoryear{{Foreman-Mackey}, {Hogg}, {Lang}  \&
  {Goodman}}{{Foreman-Mackey} et~al.}{2013}]{Foreman-Mackey2013}
{Foreman-Mackey} D.,  {Hogg} D.~W.,  {Lang} D.,   {Goodman} J.,  2013, \mn@doi
  [\pasp] {10.1086/670067}, \href
  {https://ui.adsabs.harvard.edu/abs/2013PASP..125..306F} {125, 306}

\bibitem[\protect\citeauthoryear{{Gabor} \& {Bournaud}}{{Gabor} \&
  {Bournaud}}{2013}]{Gabor2013}
{Gabor} J.~M.,  {Bournaud} F.,  2013, \mn@doi [\mnras] {10.1093/mnras/stt1046},
  \href {https://ui.adsabs.harvard.edu/abs/2013MNRAS.434..606G} {434, 606}

\bibitem[\protect\citeauthoryear{{Goodman}}{{Goodman}}{2003}]{Goodman2003}
{Goodman} J.,  2003, \mn@doi [\mnras] {10.1046/j.1365-8711.2003.06241.x}, \href
  {https://ui.adsabs.harvard.edu/abs/2003MNRAS.339..937G} {339, 937}

\bibitem[\protect\citeauthoryear{{Green} et~al.,}{{Green}
  et~al.}{2012}]{Green2012}
{Green} J.~C.,  et~al., 2012, \mn@doi [\apj] {10.1088/0004-637X/744/1/60},
  \href {https://ui.adsabs.harvard.edu/abs/2012ApJ...744...60G} {744, 60}

\bibitem[\protect\citeauthoryear{{Habouzit} et~al.,}{{Habouzit}
  et~al.}{2019}]{Habouzit2019}
{Habouzit} M.,  et~al., 2019, \mn@doi [\mnras] {10.1093/mnras/stz102}, \href
  {https://ui.adsabs.harvard.edu/abs/2019MNRAS.484.4413H} {484, 4413}

\bibitem[\protect\citeauthoryear{{Haiman} \& {Hui}}{{Haiman} \&
  {Hui}}{2001}]{Haiman2001}
{Haiman} Z.,  {Hui} L.,  2001, \mn@doi [\apj] {10.1086/318330}, \href
  {https://ui.adsabs.harvard.edu/abs/2001ApJ...547...27H} {547, 27}

\bibitem[\protect\citeauthoryear{{He} et~al.,}{{He} et~al.}{2018}]{He2018}
{He} W.,  et~al., 2018, \mn@doi [\pasj] {10.1093/pasj/psx129}, \href
  {https://ui.adsabs.harvard.edu/abs/2018PASJ...70S..33H} {70, S33}

\bibitem[\protect\citeauthoryear{{Hennawi} \& {Prochaska}}{{Hennawi} \&
  {Prochaska}}{2007}]{Hennawi2007}
{Hennawi} J.~F.,  {Prochaska} J.~X.,  2007, \mn@doi [\apj] {10.1086/509770},
  \href {https://ui.adsabs.harvard.edu/abs/2007ApJ...655..735H} {655, 735}

\bibitem[\protect\citeauthoryear{{Hogan}, {Anderson}  \& {Rugers}}{{Hogan}
  et~al.}{1997}]{Hogan1997}
{Hogan} C.~J.,  {Anderson} S.~F.,   {Rugers} M.~H.,  1997, \mn@doi [\aj]
  {10.1086/118366}, \href
  {https://ui.adsabs.harvard.edu/abs/1997AJ....113.1495H} {113, 1495}

\bibitem[\protect\citeauthoryear{{Hopkins} \& {Hernquist}}{{Hopkins} \&
  {Hernquist}}{2009}]{Hopkins2009}
{Hopkins} P.~F.,  {Hernquist} L.,  2009, \mn@doi [\apj]
  {10.1088/0004-637X/698/2/1550}, \href
  {https://ui.adsabs.harvard.edu/abs/2009ApJ...698.1550H} {698, 1550}

\bibitem[\protect\citeauthoryear{{Hopkins} \& {Quataert}}{{Hopkins} \&
  {Quataert}}{2010}]{Hopkins2010}
{Hopkins} P.~F.,  {Quataert} E.,  2010, \mn@doi [\mnras]
  {10.1111/j.1365-2966.2010.17064.x}, \href
  {https://ui.adsabs.harvard.edu/abs/2010MNRAS.407.1529H} {407, 1529}

\bibitem[\protect\citeauthoryear{{Hopkins} \& {Quataert}}{{Hopkins} \&
  {Quataert}}{2011}]{Hopkins2011a}
{Hopkins} P.~F.,  {Quataert} E.,  2011, \mn@doi [\mnras]
  {10.1111/j.1365-2966.2011.18542.x}, \href
  {https://ui.adsabs.harvard.edu/abs/2011MNRAS.415.1027H} {415, 1027}

\bibitem[\protect\citeauthoryear{{Hopkins}, {Hernquist}, {Martini}, {Cox},
  {Robertson}, {Di Matteo}  \& {Springel}}{{Hopkins}
  et~al.}{2005a}]{Hopkins2005a}
{Hopkins} P.~F.,  {Hernquist} L.,  {Martini} P.,  {Cox} T.~J.,  {Robertson} B.,
   {Di Matteo} T.,   {Springel} V.,  2005a, \mn@doi [\apjl] {10.1086/431146},
  \href {https://ui.adsabs.harvard.edu/abs/2005ApJ...625L..71H} {625, L71}

\bibitem[\protect\citeauthoryear{{Hopkins}, {Hernquist}, {Cox}, {Di Matteo},
  {Martini}, {Robertson}  \& {Springel}}{{Hopkins}
  et~al.}{2005b}]{Hopkins2005c}
{Hopkins} P.~F.,  {Hernquist} L.,  {Cox} T.~J.,  {Di Matteo} T.,  {Martini} P.,
   {Robertson} B.,   {Springel} V.,  2005b, \mn@doi [\apj] {10.1086/432438},
  \href {https://ui.adsabs.harvard.edu/abs/2005ApJ...630..705H} {630, 705}

\bibitem[\protect\citeauthoryear{{Hopkins}, {Hernquist}, {Cox}, {Di Matteo},
  {Robertson}  \& {Springel}}{{Hopkins} et~al.}{2006}]{Hopkins2006}
{Hopkins} P.~F.,  {Hernquist} L.,  {Cox} T.~J.,  {Di Matteo} T.,  {Robertson}
  B.,   {Springel} V.,  2006, \mn@doi [\apjs] {10.1086/499298}, \href
  {https://ui.adsabs.harvard.edu/abs/2006ApJS..163....1H} {163, 1}

\bibitem[\protect\citeauthoryear{{Hopkins}, {Hernquist}, {Cox}  \&
  {Kere{\v{s}}}}{{Hopkins} et~al.}{2008}]{Hopkins2008}
{Hopkins} P.~F.,  {Hernquist} L.,  {Cox} T.~J.,   {Kere{\v{s}}} D.,  2008,
  \mn@doi [\apjs] {10.1086/524362}, \href
  {https://ui.adsabs.harvard.edu/abs/2008ApJS..175..356H} {175, 356}

\bibitem[\protect\citeauthoryear{{Hui} \& {Gnedin}}{{Hui} \&
  {Gnedin}}{1997}]{Hui1997}
{Hui} L.,  {Gnedin} N.~Y.,  1997, \mn@doi [\mnras] {10.1093/mnras/292.1.27},
  \href {https://ui.adsabs.harvard.edu/abs/1997MNRAS.292...27H} {292, 27}

\bibitem[\protect\citeauthoryear{{Inayoshi}, {Visbal}  \& {Haiman}}{{Inayoshi}
  et~al.}{2019}]{Inayoshi2019}
{Inayoshi} K.,  {Visbal} E.,   {Haiman} Z.,  2019, arXiv e-prints, \href
  {https://ui.adsabs.harvard.edu/abs/2019arXiv191105791I} {p. arXiv:1911.05791}

\bibitem[\protect\citeauthoryear{{Khrykin}, {Hennawi}, {McQuinn}  \&
  {Worseck}}{{Khrykin} et~al.}{2016}]{Khrykin2016}
{Khrykin} I.~S.,  {Hennawi} J.~F.,  {McQuinn} M.,   {Worseck} G.,  2016,
  \mn@doi [\apj] {10.3847/0004-637X/824/2/133}, \href
  {http://adsabs.harvard.edu/abs/2016ApJ...824..133K} {824, 133}

\bibitem[\protect\citeauthoryear{{Khrykin}, {Hennawi}  \& {McQuinn}}{{Khrykin}
  et~al.}{2017}]{Khrykin2017}
{Khrykin} I.~S.,  {Hennawi} J.~F.,   {McQuinn} M.,  2017, \mn@doi [\apj]
  {10.3847/1538-4357/aa6621}, \href
  {http://adsabs.harvard.edu/abs/2017ApJ...838...96K} {838, 96}

\bibitem[\protect\citeauthoryear{{Khrykin}, {Hennawi}  \& {Worseck}}{{Khrykin}
  et~al.}{2019}]{Khrykin2019}
{Khrykin} I.~S.,  {Hennawi} J.~F.,   {Worseck} G.,  2019, \mn@doi [\mnras]
  {10.1093/mnras/stz135}, \href
  {http://adsabs.harvard.edu/abs/2019MNRAS.484.3897K} {484, 3897}

\bibitem[\protect\citeauthoryear{{Kormendy} \& {Ho}}{{Kormendy} \&
  {Ho}}{2013}]{Kormendy2013}
{Kormendy} J.,  {Ho} L.~C.,  2013, \mn@doi [\araa]
  {10.1146/annurev-astro-082708-101811}, \href
  {https://ui.adsabs.harvard.edu/abs/2013ARA&A..51..511K} {51, 511}

\bibitem[\protect\citeauthoryear{{Kulkarni}, {Worseck}  \&
  {Hennawi}}{{Kulkarni} et~al.}{2019}]{Kulkarni2019}
{Kulkarni} G.,  {Worseck} G.,   {Hennawi} J.~F.,  2019, \mn@doi [\mnras]
  {10.1093/mnras/stz1493}, \href
  {https://ui.adsabs.harvard.edu/abs/2019MNRAS.488.1035K} {488, 1035}

\bibitem[\protect\citeauthoryear{{La Plante}, {Trac}, {Croft}  \& {Cen}}{{La
  Plante} et~al.}{2017}]{LaPlante2017}
{La Plante} P.,  {Trac} H.,  {Croft} R.,   {Cen} R.,  2017, \mn@doi [\apj]
  {10.3847/1538-4357/aa7136}, \href
  {https://ui.adsabs.harvard.edu/abs/2017ApJ...841...87L} {841, 87}

\bibitem[\protect\citeauthoryear{{Makan}, {Worseck}, {Davies}, {Hennawi},
  {Prochaska}  \& {Richter}}{{Makan} et~al.}{2020}]{Makan2020}
{Makan} K.,  {Worseck} G.,  {Davies} F.~B.,  {Hennawi} J.~F.,  {Prochaska}
  J.~X.,   {Richter} P.,  2020, arXiv e-prints, \href
  {https://ui.adsabs.harvard.edu/abs/2020arXiv201207876M} {p. arXiv:2012.07876}

\bibitem[\protect\citeauthoryear{{Martini}}{{Martini}}{2004}]{Martini2004}
{Martini} P.,  2004, in {Ho} L.~C.,  ed., Coevolution of Black Holes and
  Galaxies. p.~169 (\mn@eprint {arXiv} {astro-ph/0304009})

\bibitem[\protect\citeauthoryear{{Martini} \& {Weinberg}}{{Martini} \&
  {Weinberg}}{2001}]{Martini2001}
{Martini} P.,  {Weinberg} D.~H.,  2001, \mn@doi [\apj] {10.1086/318331}, \href
  {https://ui.adsabs.harvard.edu/abs/2001ApJ...547...12M} {547, 12}

\bibitem[\protect\citeauthoryear{{Mazzucchelli} et~al.,}{{Mazzucchelli}
  et~al.}{2017}]{Mazzucchelli2017}
{Mazzucchelli} C.,  et~al., 2017, \mn@doi [\apj] {10.3847/1538-4357/aa9185},
  \href {https://ui.adsabs.harvard.edu/abs/2017ApJ...849...91M} {849, 91}

\bibitem[\protect\citeauthoryear{{McQuinn}, {Lidz}, {Zaldarriaga}, {Hernquist},
  {Hopkins}, {Dutta}  \& {Faucher-Gigu{\`e}re}}{{McQuinn}
  et~al.}{2009}]{McQuinn2009}
{McQuinn} M.,  {Lidz} A.,  {Zaldarriaga} M.,  {Hernquist} L.,  {Hopkins} P.~F.,
   {Dutta} S.,   {Faucher-Gigu{\`e}re} C.-A.,  2009, \mn@doi [\apj]
  {10.1088/0004-637X/694/2/842}, \href
  {https://ui.adsabs.harvard.edu/abs/2009ApJ...694..842M} {694, 842}

\bibitem[\protect\citeauthoryear{{Mellema}, {Iliev}, {Alvarez}  \&
  {Shapiro}}{{Mellema} et~al.}{2006}]{mellema2006}
{Mellema} G.,  {Iliev} I.~T.,  {Alvarez} M.~A.,   {Shapiro} P.~R.,  2006,
  \mn@doi [NA] {10.1016/j.newast.2005.09.004}, \href
  {http://adsabs.harvard.edu/abs/2006NewA...11..374M} {11, 374}

\bibitem[\protect\citeauthoryear{{Merloni} et~al.,}{{Merloni}
  et~al.}{2014}]{Merloni2014}
{Merloni} A.,  et~al., 2014, \mn@doi [\mnras] {10.1093/mnras/stt2149}, \href
  {https://ui.adsabs.harvard.edu/abs/2014MNRAS.437.3550M} {437, 3550}

\bibitem[\protect\citeauthoryear{{Morey}, {Eilers}  et~al.}{{Morey}
  et~al.}{prep}]{Morey2021}
{Morey} K.,  {Eilers} A.~C.,   et~al., in prep

\bibitem[\protect\citeauthoryear{{Mortlock} et~al.,}{{Mortlock}
  et~al.}{2011}]{Mortlock2011}
{Mortlock} D.~J.,  et~al., 2011, \mn@doi [\nat] {10.1038/nature10159}, \href
  {https://ui.adsabs.harvard.edu/abs/2011Natur.474..616M} {474, 616}

\bibitem[\protect\citeauthoryear{{Novak}, {Ostriker}  \& {Ciotti}}{{Novak}
  et~al.}{2011}]{Novak2011}
{Novak} G.~S.,  {Ostriker} J.~P.,   {Ciotti} L.,  2011, \mn@doi [\apj]
  {10.1088/0004-637X/737/1/26}, \href
  {https://ui.adsabs.harvard.edu/abs/2011ApJ...737...26N} {737, 26}

\bibitem[\protect\citeauthoryear{{Planck Collaboration} et~al.,}{{Planck
  Collaboration} et~al.}{2018}]{Planck2018}
{Planck Collaboration} et~al., 2018, arXiv e-prints, \href
  {https://ui.adsabs.harvard.edu/abs/2018arXiv180706209P} {p. arXiv:1807.06209}

\bibitem[\protect\citeauthoryear{{Polletta}, {Weedman}, {H{\"o}nig},
  {Lonsdale}, {Smith}  \& {Houck}}{{Polletta} et~al.}{2008}]{Poletta2008}
{Polletta} M.,  {Weedman} D.,  {H{\"o}nig} S.,  {Lonsdale} C.~J.,  {Smith}
  H.~E.,   {Houck} J.,  2008, \mn@doi [\apj] {10.1086/524343}, \href
  {https://ui.adsabs.harvard.edu/abs/2008ApJ...675..960P} {675, 960}

\bibitem[\protect\citeauthoryear{{Puchwein}, {Haardt}, {Haehnelt}  \&
  {Madau}}{{Puchwein} et~al.}{2019}]{Puchwein2019}
{Puchwein} E.,  {Haardt} F.,  {Haehnelt} M.~G.,   {Madau} P.,  2019, \mn@doi
  [\mnras] {10.1093/mnras/stz222}, \href
  {https://ui.adsabs.harvard.edu/abs/2019MNRAS.485...47P} {485, 47}

\bibitem[\protect\citeauthoryear{{Regan}, {Downes}, {Volonteri}, {Beckmann},
  {Lupi}, {Trebitsch}  \& {Dubois}}{{Regan} et~al.}{2019}]{Regan2019}
{Regan} J.~A.,  {Downes} T.~P.,  {Volonteri} M.,  {Beckmann} R.,  {Lupi} A.,
  {Trebitsch} M.,   {Dubois} Y.,  2019, \mn@doi [\mnras]
  {10.1093/mnras/stz1045}, \href
  {https://ui.adsabs.harvard.edu/abs/2019MNRAS.486.3892R} {486, 3892}

\bibitem[\protect\citeauthoryear{{Salpeter}}{{Salpeter}}{1964}]{Salpeter1964}
{Salpeter} E.~E.,  1964, \mn@doi [\apj] {10.1086/147973}, \href
  {https://ui.adsabs.harvard.edu/abs/1964ApJ...140..796S} {140, 796}

\bibitem[\protect\citeauthoryear{{Schawinski}, {Koss}, {Berney}  \&
  {Sartori}}{{Schawinski} et~al.}{2015}]{Schawinski2015}
{Schawinski} K.,  {Koss} M.,  {Berney} S.,   {Sartori} L.~F.,  2015, \mn@doi
  [\mnras] {10.1093/mnras/stv1136}, \href
  {https://ui.adsabs.harvard.edu/abs/2015MNRAS.451.2517S} {451, 2517}

\bibitem[\protect\citeauthoryear{{Schmidt}, {Worseck}, {Hennawi}, {Prochaska}
  \& {Crighton}}{{Schmidt} et~al.}{2017}]{Schmidt2017}
{Schmidt} T.~M.,  {Worseck} G.,  {Hennawi} J.~F.,  {Prochaska} J.~X.,
  {Crighton} N. H.~M.,  2017, \mn@doi [\apj] {10.3847/1538-4357/aa83ac}, \href
  {https://ui.adsabs.harvard.edu/abs/2017ApJ...847...81S} {847, 81}

\bibitem[\protect\citeauthoryear{{Schmidt}, {Hennawi}, {Worseck}, {Davies},
  {Luki{\'c}}  \& {O{\~n}orbe}}{{Schmidt} et~al.}{2018}]{Schmidt2018}
{Schmidt} T.~M.,  {Hennawi} J.~F.,  {Worseck} G.,  {Davies} F.~B.,  {Luki{\'c}}
  Z.,   {O{\~n}orbe} J.,  2018, \mn@doi [\apj] {10.3847/1538-4357/aac8e4},
  \href {https://ui.adsabs.harvard.edu/abs/2018ApJ...861..122S} {861, 122}

\bibitem[\protect\citeauthoryear{{Shankar}, {Weinberg}  \&
  {Miralda-Escud{\'e}}}{{Shankar} et~al.}{2009}]{Shankar2009}
{Shankar} F.,  {Weinberg} D.~H.,   {Miralda-Escud{\'e}} J.,  2009, \mn@doi
  [\apj] {10.1088/0004-637X/690/1/20}, \href
  {https://ui.adsabs.harvard.edu/abs/2009ApJ...690...20S} {690, 20}

\bibitem[\protect\citeauthoryear{{Shankar}, {Weinberg}  \& {Shen}}{{Shankar}
  et~al.}{2010}]{Shankar2010}
{Shankar} F.,  {Weinberg} D.~H.,   {Shen} Y.,  2010, \mn@doi [\mnras]
  {10.1111/j.1365-2966.2010.16801.x}, \href
  {https://ui.adsabs.harvard.edu/abs/2010MNRAS.406.1959S} {406, 1959}

\bibitem[\protect\citeauthoryear{{Shen} et~al.,}{{Shen}
  et~al.}{2007}]{Shen2007}
{Shen} Y.,  et~al., 2007, \mn@doi [\aj] {10.1086/513517}, \href
  {https://ui.adsabs.harvard.edu/abs/2007AJ....133.2222S} {133, 2222}

\bibitem[\protect\citeauthoryear{{Shen} et~al.,}{{Shen}
  et~al.}{2009}]{Shen2009}
{Shen} Y.,  et~al., 2009, \mn@doi [\apj] {10.1088/0004-637X/697/2/1656}, \href
  {https://ui.adsabs.harvard.edu/abs/2009ApJ...697.1656S} {697, 1656}

\bibitem[\protect\citeauthoryear{{Shen} et~al.,}{{Shen}
  et~al.}{2019}]{Shen2019}
{Shen} Y.,  et~al., 2019, \mn@doi [\apj] {10.3847/1538-4357/ab03d9}, \href
  {https://ui.adsabs.harvard.edu/abs/2019ApJ...873...35S} {873, 35}

\bibitem[\protect\citeauthoryear{{Smith}, {Bromm}  \& {Loeb}}{{Smith}
  et~al.}{2017}]{Smith2017}
{Smith} A.,  {Bromm} V.,   {Loeb} A.,  2017, \mn@doi [Astronomy and Geophysics]
  {10.1093/astrogeo/atx099}, \href
  {https://ui.adsabs.harvard.edu/abs/2017A&G....58c3.22S} {58, 3.22}

\bibitem[\protect\citeauthoryear{{Soltan}}{{Soltan}}{1982}]{Soltan1982}
{Soltan} A.,  1982, \mn@doi [\mnras] {10.1093/mnras/200.1.115}, \href
  {https://ui.adsabs.harvard.edu/abs/1982MNRAS.200..115S} {200, 115}

\bibitem[\protect\citeauthoryear{{Springel}}{{Springel}}{2005}]{springel2005}
{Springel} V.,  2005, \mn@doi [\mnras] {10.1111/j.1365-2966.2005.09655.x},
  \href {https://ui.adsabs.harvard.edu/abs/2005MNRAS.364.1105S} {364, 1105}

\bibitem[\protect\citeauthoryear{{Springel}, {Di Matteo}  \&
  {Hernquist}}{{Springel} et~al.}{2005}]{Springel2005a}
{Springel} V.,  {Di Matteo} T.,   {Hernquist} L.,  2005, \mn@doi [\mnras]
  {10.1111/j.1365-2966.2005.09238.x}, \href
  {https://ui.adsabs.harvard.edu/abs/2005MNRAS.361..776S} {361, 776}

\bibitem[\protect\citeauthoryear{{Steinborn}, {Hirschmann}, {Dolag}, {Shankar},
  {Juneau}, {Krumpe}, {Remus}  \& {Teklu}}{{Steinborn}
  et~al.}{2018}]{Steinborn2018}
{Steinborn} L.~K.,  {Hirschmann} M.,  {Dolag} K.,  {Shankar} F.,  {Juneau} S.,
  {Krumpe} M.,  {Remus} R.-S.,   {Teklu} A.~F.,  2018, \mn@doi [\mnras]
  {10.1093/mnras/sty2288}, \href
  {https://ui.adsabs.harvard.edu/abs/2018MNRAS.481..341S} {481, 341}

\bibitem[\protect\citeauthoryear{{Syphers} \& {Shull}}{{Syphers} \&
  {Shull}}{2014}]{Syphers2014}
{Syphers} D.,  {Shull} J.~M.,  2014, \mn@doi [\apj]
  {10.1088/0004-637X/784/1/42}, \href
  {https://ui.adsabs.harvard.edu/abs/2014ApJ...784...42S} {784, 42}

\bibitem[\protect\citeauthoryear{{Thorne}}{{Thorne}}{1974}]{Thorne1974}
{Thorne} K.~S.,  1974, \mn@doi [\apj] {10.1086/152991}, \href
  {https://ui.adsabs.harvard.edu/abs/1974ApJ...191..507T} {191, 507}

\bibitem[\protect\citeauthoryear{{Ueda}, {Akiyama}, {Hasinger}, {Miyaji}  \&
  {Watson}}{{Ueda} et~al.}{2014}]{Ueda2014}
{Ueda} Y.,  {Akiyama} M.,  {Hasinger} G.,  {Miyaji} T.,   {Watson} M.~G.,
  2014, \mn@doi [\apj] {10.1088/0004-637X/786/2/104}, \href
  {https://ui.adsabs.harvard.edu/abs/2014ApJ...786..104U} {786, 104}

\bibitem[\protect\citeauthoryear{{Venemans} et~al.,}{{Venemans}
  et~al.}{2013}]{Venemans2013}
{Venemans} B.~P.,  et~al., 2013, \mn@doi [\apj] {10.1088/0004-637X/779/1/24},
  \href {https://ui.adsabs.harvard.edu/abs/2013ApJ...779...24V} {779, 24}

\bibitem[\protect\citeauthoryear{{Vito} et~al.,}{{Vito}
  et~al.}{2018}]{Vito2018}
{Vito} F.,  et~al., 2018, \mn@doi [\mnras] {10.1093/mnras/stx2486}, \href
  {https://ui.adsabs.harvard.edu/abs/2018MNRAS.473.2378V} {473, 2378}

\bibitem[\protect\citeauthoryear{{Volonteri}}{{Volonteri}}{2010}]{Volonteri2010}
{Volonteri} M.,  2010, \mn@doi [\aapr] {10.1007/s00159-010-0029-x}, \href
  {https://ui.adsabs.harvard.edu/abs/2010A&ARv..18..279V} {18, 279}

\bibitem[\protect\citeauthoryear{{Volonteri}}{{Volonteri}}{2012}]{Volonteri2012}
{Volonteri} M.,  2012, \mn@doi [Science] {10.1126/science.1220843}, \href
  {https://ui.adsabs.harvard.edu/abs/2012Sci...337..544V} {337, 544}

\bibitem[\protect\citeauthoryear{{Wang} et~al.,}{{Wang}
  et~al.}{2020}]{Wang2020}
{Wang} F.,  et~al., 2020, \mn@doi [\apj] {10.3847/1538-4357/ab8c45}, \href
  {https://ui.adsabs.harvard.edu/abs/2020ApJ...896...23W} {896, 23}

\bibitem[\protect\citeauthoryear{{Wang} et~al.,}{{Wang}
  et~al.}{2021}]{Wang2021}
{Wang} F.,  et~al., 2021, arXiv e-prints, \href
  {https://ui.adsabs.harvard.edu/abs/2021arXiv210103179W} {p. arXiv:2101.03179}

\bibitem[\protect\citeauthoryear{{White}, {Martini}  \& {Cohn}}{{White}
  et~al.}{2008}]{White2008}
{White} M.,  {Martini} P.,   {Cohn} J.~D.,  2008, \mn@doi [\mnras]
  {10.1111/j.1365-2966.2008.13817.x}, \href
  {https://ui.adsabs.harvard.edu/abs/2008MNRAS.390.1179W} {390, 1179}

\bibitem[\protect\citeauthoryear{{White} et~al.,}{{White}
  et~al.}{2012}]{White2012}
{White} M.,  et~al., 2012, \mn@doi [\mnras] {10.1111/j.1365-2966.2012.21251.x},
  \href {https://ui.adsabs.harvard.edu/abs/2012MNRAS.424..933W} {424, 933}

\bibitem[\protect\citeauthoryear{{Worseck}, {Davies}, {Hennawi}  \&
  {Prochaska}}{{Worseck} et~al.}{2019}]{Worseck2019}
{Worseck} G.,  {Davies} F.~B.,  {Hennawi} J.~F.,   {Prochaska} J.~X.,  2019,
  \mn@doi [\apj] {10.3847/1538-4357/ab0fa1}, \href
  {https://ui.adsabs.harvard.edu/abs/2019ApJ...875..111W} {875, 111}

\bibitem[\protect\citeauthoryear{{Worseck}, {Khrykin}, {Hennawi}, {Prochaska}
  \& {Farina}}{{Worseck} et~al.}{2021}]{Worseck2021}
{Worseck} G.,  {Khrykin} I.~S.,  {Hennawi} J.~F.,  {Prochaska} J.~X.,
  {Farina} E.~P.,  2021, arXiv e-prints, \href
  {https://ui.adsabs.harvard.edu/abs/2021arXiv210101196W} {p. arXiv:2101.01196}

\bibitem[\protect\citeauthoryear{{Wu} et~al.,}{{Wu} et~al.}{2015}]{Wu2015}
{Wu} X.-B.,  et~al., 2015, in IAU General Assembly. p. 2251223

\bibitem[\protect\citeauthoryear{{Yang} et~al.,}{{Yang}
  et~al.}{2020a}]{Yang2020a}
{Yang} J.,  et~al., 2020a, arXiv e-prints, \href
  {https://ui.adsabs.harvard.edu/abs/2020arXiv200613452Y} {p. arXiv:2006.13452}

\bibitem[\protect\citeauthoryear{{Yang} et~al.,}{{Yang}
  et~al.}{2020b}]{Yang2020b}
{Yang} J.,  et~al., 2020b, \mn@doi [\apj] {10.3847/1538-4357/abbc1b}, \href
  {https://ui.adsabs.harvard.edu/abs/2020ApJ...904...26Y} {904, 26}

\bibitem[\protect\citeauthoryear{{Yu} \& {Tremaine}}{{Yu} \&
  {Tremaine}}{2002}]{Yu2002}
{Yu} Q.,  {Tremaine} S.,  2002, \mn@doi [\mnras]
  {10.1046/j.1365-8711.2002.05532.x}, \href
  {https://ui.adsabs.harvard.edu/abs/2002MNRAS.335..965Y} {335, 965}

\bibitem[\protect\citeauthoryear{{Zheng}, {Syphers}, {Meiksin}, {Kriss},
  {Schneider}, {York}  \& {Anderson}}{{Zheng} et~al.}{2015}]{Zheng2015}
{Zheng} W.,  {Syphers} D.,  {Meiksin} A.,  {Kriss} G.~A.,  {Schneider} D.~P.,
  {York} D.~G.,   {Anderson} S.~F.,  2015, \mn@doi [\apj]
  {10.1088/0004-637X/806/1/142}, \href
  {https://ui.adsabs.harvard.edu/abs/2015ApJ...806..142Z} {806, 142}

\makeatother
\end{thebibliography}




\appendix

\section{Analytical Solution for Distribution of on-times}
\label{sec:app}

In what follows we derive the analytical solution for the probability density function of the logarithm of the quasar on-times $t_{\rm on}$, $p\left( {\rm log}_{10}t_{\rm on} \right)$. Given the definition of the on-times, they follow the uniform distribution (see Section~\ref{sec:data}), and the conditional probability of $t_{\rm on}$ given the quasar lifetime $t_{\rm Q}$ is then can be written as 
\begin{equation}
\label{eq:pton}
    p \left( t_{\rm on} | t_{\rm Q} \right) = \begin{cases}
               \frac{1}{t_{\rm Q}},\ 0 \leq t_{\rm on} \leq t_{\rm Q}\\
               0,\ t_{\rm on} < 0, t_{\rm on} > t_{\rm Q}.
            \end{cases}
\end{equation}

On the other hand, we assumed a base-10 log-normal distribution of quasar lifetimes $t_{\rm Q}$, for which the probability density function is given by eq.~(\ref{eq:qld})

\begin{equation}
\label{eq:ptq}
    p\left( t_{\rm Q} \right) = \frac{1}{t_{\rm Q}} \frac{ {\rm log_{10}e} }{ \sigma \sqrt{2\pi} } {\rm exp}\left[ - \frac{\left( \log_{10}\left(  t_{\rm Q} / {\rm Myr} \right) - \mu \right)^2}{ 2 \sigma^2 } \right],
\end{equation}
where $\mu = \langle {\rm log}_{10}\left( t_{\rm Q} / {\rm Myr} \right) \rangle$ and $\sigma = \sigma_{\rm log_{10}t_{\rm Q}}$ are mean and standard deviation of QLD distribution. The rule of total probability dictates that

\begin{equation}
\label{eq:rtp}
    p\left( t_{\rm on} \right) = \int p \left( t_{\rm on} | t_{\rm Q} \right) p\left( t_{\rm Q} \right) {\rm d}t_{\rm Q}.
\end{equation}

Taking into account eq.~(\ref{eq:pton}) and eq.~(\ref{eq:ptq}),  eq.~(\ref{eq:rtp}) becomes

\begin{equation}
\label{eq:pton2}
    p\left( t_{\rm on} \right) = \int_{t_{\rm on}}^{\infty} \frac{1}{t_{\rm Q}} \cdot \frac{{\rm log_{10}e}}{t_{\rm Q}\sigma \sqrt{2\pi}} {\rm exp}\left[ - \frac{\left( \log_{10}\left(  t_{\rm Q} / {\rm Myr} \right) - \mu \right)^2}{ 2 \sigma^2 } \right] {\rm d}t_{\rm Q}.
\end{equation}

Assuming $u = {\rm log_{10}}t_{\rm Q}$ and changing the variables, we can re-write eq.~(\ref{eq:pton2}) as follows

\begin{equation}
\label{eq:pton3}
    p\left( t_{\rm on} \right) = \frac{1}{\sigma \sqrt{2\pi}}  \int_{{\rm log_{10}}t_{\rm on}}^{\infty} 10^{-u}\cdot {\rm exp}\left[ - \frac{\left(u - \mu \right)^2}{ 2 \sigma^2 } \right] {\rm d}u.
\end{equation}
This integral has analytical solution given by

\begin{equation}
\label{eq:t_on_final}
    \begin{aligned}
    p\left( t_{\rm on} \right) = \frac{2^{-\mu-1/2} 5^{-\mu}}{\sqrt{2}}{\rm exp}\left[ \frac{1}{2}\sigma^2 {\rm ln^2}10 \right] \cdot \\
    {\rm Erfc}\left[ \frac{{\rm log_{10}}\left(t_{\rm on}/{\rm Myr}\right) - \mu + \sigma^2{\rm ln}10}{\sigma\sqrt{2}} \right],
    \end{aligned}
\end{equation}
where Erfc is the complementary error function. Finally, in order to find the desired probability $p\left( {\rm log_{10}} \left( t_{\rm on} / {\rm Myr} \right) \right)$, we apply the transformation of the variables and find that

\begin{equation}
\label{eq:logt_on}
    p\left( {\rm log_{10}} \left( t_{\rm on} / {\rm Myr} \right) \right) = {\rm ln} 10~t_{\rm on} \cdot p\left( t_{\rm on} \right).
\end{equation}

We illustrate this analytical solution for the $p\left( {\rm log_{10}} \left( t_{\rm on} / {\rm Myr} \right) \right)$ given eq.~(\ref{eq:t_on_final}) and eq.~(\ref{eq:logt_on}) in the top panel of Fig.~\ref{fig:violin}. 

\bsp	
\label{lastpage}
\end{document}